\documentclass[12pt]{article}
\usepackage{amsmath,amssymb}
\numberwithin{equation}{section}
\title{Chiral Algebras of $(0,2)$ Sigma Models:\\
Beyond Perturbation Theory \\[12pt]}

\author{\normalsize Meng-Chwan~Tan\footnote{On leave of absence from the National University of Singapore.} \\
\normalsize\it
School of Natural Sciences, Institute for Advanced Study \\
\normalsize\it
Princeton, NJ 08540, USA  \\[2pt]
\normalsize and \\[2pt]
\normalsize Junya Yagi \\
\normalsize\it
Department of Physics and Astronomy, Rutgers University \\
\normalsize\it Piscataway, NJ 08855, USA}

\date{}

\makeatletter
\renewcommand\section{\@startsection {section}{1}{\z@}%
                                   {-3.5ex \@plus -1ex \@minus -.2ex}%
                                   {2.3ex \@plus.2ex}%
                                   {\normalfont\Large\bfseries\boldmath}}
\renewcommand\subsection{\@startsection{subsection}{2}{\z@}%
                                     {-3.25ex\@plus -1ex \@minus -.2ex}%
                                     {1.5ex \@plus .2ex}%
                                     {\normalfont\large\bfseries\boldmath}}
\renewcommand\subsubsection{\@startsection{subsubsection}{3}{\z@}%
                                     {-3.25ex\@plus -1ex \@minus -.2ex}%
                                     {1.5ex \@plus .2ex}%
                                     {\normalfont\normalsize\bfseries\boldmath}}
\renewcommand\paragraph{\@startsection{paragraph}{4}{\z@}%
                                    {3.25ex \@plus1ex \@minus.2ex}%
                                    {-1em}%
                                    {\normalfont\normalsize\bfseries\boldmath}}
\renewcommand\subparagraph{\@startsection{subparagraph}{5}{\parindent}%
                                       {3.25ex \@plus1ex \@minus .2ex}%
                                       {-1em}%
                                      {\normalfont\normalsize\bfseries\boldmath}}
\makeatother

\newcommand{\del}{\partial}
\newcommand{\delb}{{\bar\partial}}
\newcommand{\Det}{\mathop{\mathrm{Det}}\nolimits}
\newcommand{\TXb}{\overline{TX}}
\newcommand{\CP}{\mathbb{CP}}
\newcommand{\set}[1]{\{#1\}}
\newcommand{\ip}[2]{\langle #1, #2\rangle}
\newcommand{\ket}[1]{|#1\rangle}
\newcommand{\bra}[1]{\langle #1|}
\newcommand{\blank}{\mathord{\,\cdot\,}}
\newcommand{\vev}[1]{\langle #1 \rangle}
\newcommand{\bigvev}[1]{\bigl\langle #1 \bigr\rangle}
\newcommand{\Bigvev}[1]{\Bigl\langle #1 \Bigr\rangle}
\newcommand{\CNO}[1]{\mathopen{:}#1\mathclose{:}} 

\newcommand{\Map}{\mathop{\mathrm{Map}}\nolimits}
\newcommand{\Sym}{\mathop{\mathrm{Sym}}\nolimits}
\newcommand{\Tr}{\mathop{\mathrm{Tr}}\nolimits}
\newcommand{\longto}{\longrightarrow}
\newcommand{\iso}{\cong}

\newcommand{\Z}{\mathbb{Z}}

\newcommand{\R}{\mathbb{R}}
\newcommand{\C}{\mathbb{C}}

\let\nc\newcommand
\let\ovl\overline
\let\td\tilde
\let\wtd\widetilde
\let\wht\widehat
\let\mcl\mathcal

\newcommand{\ab}{{\bar{a}}}
\newcommand{\bb}{{\bar{b}}}
\newcommand{\cb}{{\bar{c}}}
\newcommand{\db}{{\bar{d}}}

\newcommand{\hb}{{\bar{h}}}
\newcommand{\ib}{{\bar{\imath}}}
\newcommand{\jb}{{\bar{\jmath}}}
\newcommand{\kb}{{\bar{k}}}
\newcommand{\lb}{{\bar{l}}}

\newcommand{\nb}{{\bar{n}}}

\newcommand{\ub}{{\bar{u}}}

\newcommand{\wb}{{\bar{w}}}

\newcommand{\zb}{{\bar{z}}}

\nc{\Ab}{{\ovl{A}}} \nc{\At}{{\wtd{A}}} \nc{\Ah}{{\wht{A}}} \nc{\Bb}{{\ovl{B}}} \nc{\Bt}{{\wtd{B}}} \nc{\Bh}{{\wht{B}}} \nc{\Cb}{{\ovl{C}}}
\nc{\Ct}{{\wtd{C}}} \nc{\Ch}{{\wht{C}}} \nc{\Db}{{\ovl{D}}} \nc{\Dt}{{\wtd{D}}} \nc{\Dh}{{\wht{D}}} \nc{\Eb}{{\ovl{E}}} \nc{\Et}{{\wtd{E}}}
\nc{\Eh}{{\wht{E}}} \nc{\Fb}{{\ovl{F}}} \nc{\Ft}{{\wtd{F}}} \nc{\Fh}{{\wht{F}}} \nc{\Gb}{{\ovl{G}}} \nc{\Gt}{{\wtd{G}}} \nc{\Gh}{{\wht{G}}}
\nc{\Hb}{{\ovl{H}}} \nc{\Ht}{{\wtd{H}}} \nc{\Hh}{{\wht{H}}} \nc{\Ib}{{\ovl{I}}} \nc{\It}{{\wtd{I}}} \nc{\Ih}{{\wht{I}}} \nc{\Jb}{{\ovl{J}}}
\nc{\Jt}{{\wtd{J}}} \nc{\Jh}{{\wht{J}}} \nc{\Kb}{{\ovl{K}}} \nc{\Kt}{{\wtd{K}}} \nc{\Kh}{{\wht{K}}} \nc{\Lb}{{\ovl{L}}} \nc{\Lt}{{\wtd{L}}}
\nc{\Lh}{{\wht{L}}} \nc{\Mb}{{\ovl{M}}} \nc{\Mt}{{\wtd{M}}} \nc{\Mh}{{\wht{M}}} \nc{\Nb}{{\ovl{N}}} \nc{\Nt}{{\wtd{N}}} \nc{\Nh}{{\wht{N}}}
\nc{\Ob}{{\ovl{O}}} \nc{\Ot}{{\wtd{O}}} \nc{\Oh}{{\wht{O}}} \nc{\Pb}{{\ovl{P}}} \nc{\Pt}{{\wtd{P}}} \nc{\Ph}{{\wht{P}}} \nc{\Qb}{{\ovl{Q}}}
\nc{\Qt}{{\wtd{Q}}} \nc{\Qh}{{\wht{Q}}} \nc{\Rb}{{\ovl{R}}} \nc{\Rt}{{\wtd{R}}} \nc{\Rh}{{\wht{R}}} \nc{\Sb}{{\ovl{S}}} \nc{\St}{{\wtd{S}}}
\nc{\Sh}{{\wht{S}}} \nc{\Tb}{{\ovl{T}}} \nc{\Tt}{{\wtd{T}}} \nc{\Th}{{\wht{T}}} \nc{\Ub}{{\ovl{U}}} \nc{\Ut}{{\wtd{U}}} \nc{\Uh}{{\wht{U}}}
\nc{\Vb}{{\ovl{V}}} \nc{\Vt}{{\wtd{V}}} \nc{\Vh}{{\wht{V}}} \nc{\Wb}{{\ovl{W}}} \nc{\Wt}{{\wtd{W}}} \nc{\Wh}{{\wht{W}}} \nc{\Xb}{{\ovl{X}}}
\nc{\Xt}{{\wtd{X}}} \nc{\Xh}{{\wht{X}}} \nc{\Yb}{{\ovl{Y}}} \nc{\Yt}{{\wtd{Y}}} \nc{\Yh}{{\wht{Y}}} \nc{\Zb}{{\ovl{Z}}} \nc{\Zt}{{\wtd{Z}}}
\nc{\Zh}{{\wht{Z}}}

\nc{\CA}{{\mcl{A}}} \nc{\CAb}{{\ovl{\CA}}} \nc{\CAt}{{\wtd{\CA}}} \nc{\CAh}{{\wht{\CA}}} \nc{\CB}{{\mcl{B}}} \nc{\CBb}{{\ovl{\CB}}}
\nc{\CBt}{{\wtd{\CB}}} \nc{\CBh}{{\wht{\CB}}} \nc{\CC}{{\mcl{C}}} \nc{\CCb}{{\ovl{\CC}}} \nc{\CCt}{{\wtd{\CC}}} \nc{\CCh}{{\wht{\CC}}}
\nc{\CD}{{\mcl{D}}} \nc{\CDb}{{\ovl{\CD}}} \nc{\CDt}{{\wtd{\CD}}} \nc{\CDh}{{\wht{\CD}}} \nc{\CE}{{\mcl{E}}} \nc{\CEb}{{\ovl{\CE}}}
\nc{\CEt}{{\wtd{\CE}}} \nc{\CEh}{{\wht{\CE}}} \nc{\CF}{{\mcl{F}}} \nc{\CFb}{{\ovl{\CF}}} \nc{\CFt}{{\wtd{\CF}}} \nc{\CFh}{{\wht{\CF}}}
\nc{\CG}{{\mcl{G}}} \nc{\CGb}{{\ovl{\CG}}} \nc{\CGt}{{\wtd{\CG}}} \nc{\CGh}{{\wht{\CG}}} \nc{\CH}{{\mcl{H}}} \nc{\CHb}{{\ovl{\CH}}}
\nc{\CHt}{{\wtd{\CH}}} \nc{\CHh}{{\wht{\CH}}} \nc{\CI}{{\mcl{I}}} \nc{\CIb}{{\ovl{\CI}}} \nc{\CIt}{{\wtd{\CI}}} \nc{\CIh}{{\wht{\CI}}}
\nc{\CJ}{{\mcl{J}}} \nc{\CJb}{{\ovl{\CJ}}} \nc{\CJt}{{\wtd{\CJ}}} \nc{\CJh}{{\wht{\CJ}}} \nc{\CK}{{\mcl{K}}} \nc{\CKb}{{\ovl{\CK}}}
\nc{\CKt}{{\wtd{\CK}}} \nc{\CKh}{{\wht{\CK}}} \nc{\CL}{{\mcl{L}}} \nc{\CLb}{{\ovl{\CL}}} \nc{\CLt}{{\wtd{\CL}}} \nc{\CLh}{{\wht{\CL}}}
\nc{\CM}{{\mcl{M}}} \nc{\CMb}{{\ovl{\CM}}} \nc{\CMt}{{\wtd{\CM}}} \nc{\CMh}{{\wht{\CM}}} \nc{\CN}{{\mcl{N}}} \nc{\CNb}{{\ovl{\CN}}}
\nc{\CNt}{{\wtd{\CN}}} \nc{\CNh}{{\wht{\CN}}} \nc{\CO}{{\mcl{O}}} \nc{\COb}{{\ovl{\CO}}} \nc{\COt}{{\wtd{\CO}}} \nc{\COh}{{\wht{\CO}}}
\nc{\CQ}{{\mcl{Q}}} \nc{\CQb}{{\ovl{\CQ}}} \nc{\CQt}{{\wtd{\CQ}}} \nc{\CQh}{{\wht{\CQ}}} \nc{\CR}{{\mcl{R}}} \nc{\CRb}{{\ovl{\CR}}}
\nc{\CRt}{{\wtd{\CR}}} \nc{\CRh}{{\wht{\CR}}} \nc{\CS}{{\mcl{S}}} \nc{\CSb}{{\ovl{\CS}}} \nc{\CSt}{{\wtd{\CS}}} \nc{\CSh}{{\wht{\CS}}}
\nc{\CT}{{\mcl{T}}} \nc{\CTb}{{\ovl{\CT}}} \nc{\CTt}{{\wtd{\CT}}} \nc{\CTh}{{\wht{\CT}}} \nc{\CU}{{\mcl{U}}} \nc{\CUb}{{\ovl{\CU}}}
\nc{\CUt}{{\wtd{\CU}}} \nc{\CUh}{{\wht{\CU}}} \nc{\CV}{{\mcl{V}}} \nc{\CVb}{{\ovl{\CV}}} \nc{\CVt}{{\wtd{\CV}}} \nc{\CVh}{{\wht{\CV}}}
\nc{\CW}{{\mcl{W}}} \nc{\CWb}{{\ovl{\CW}}} \nc{\CWt}{{\wtd{\CW}}} \nc{\CWh}{{\wht{\CW}}} \nc{\CX}{{\mcl{X}}} \nc{\CXb}{{\ovl{\CX}}}
\nc{\CXt}{{\wtd{\CX}}} \nc{\CXh}{{\wht{\CX}}} \nc{\CY}{{\mcl{Y}}} \nc{\CYb}{{\ovl{\CY}}} \nc{\CYt}{{\wtd{\CY}}} \nc{\CYh}{{\wht{\CY}}}
\nc{\CZ}{{\mcl{Z}}} \nc{\CZb}{{\ovl{\CZ}}} \nc{\CZt}{{\wtd{\CZ}}} \nc{\CZh}{{\wht{\CZ}}}

\let\eps\epsilon
\let\ups\upsilon
\let\veps\varepsilon
\let\vtht\vartheta
\let\vsgm\varsigma
\let\vphi\varphi
\let\vrho\varrho

\nc{\alphab}{{\bar{\alpha}}} \nc{\alphat}{{\td{\alpha}}} \nc{\alphah}{{\hat{\alpha}}} \nc{\betab}{{\bar{\beta}}}   \nc{\betat}{{\td{\beta}}}
\nc{\betah}{{\hat{\beta}}} \nc{\gammab}{{\bar{\gamma}}} \nc{\gammat}{{\td{\gamma}}} \nc{\gammah}{{\hat{\gamma}}} \nc{\deltab}{{\bar{\delta}}}
\nc{\deltat}{{\td{\delta}}} \nc{\deltah}{{\hat{\delta}}} \nc{\epsilonb}{{\bar{\eps}}} \nc{\epsilont}{{\td{\eps}}} \nc{\epsilonh}{{\hat{\eps}}}
\nc{\vepsb}{{\bar{\veps}}}   \nc{\vepst}{{\td{\veps}}}   \nc{\vepsh}{{\hat{\veps}}} \nc{\zetab}{{\bar{\zeta}}}   \nc{\zetat}{{\td{\zeta}}}
\nc{\zetah}{{\hat{\zeta}}} \nc{\etab}{{\bar{\eta}}}     \nc{\etat}{{\td{\eta}}}     \nc{\etah}{{\hat{\eta}}} \nc{\thetab}{{\bar{\theta}}}
\nc{\thetat}{{\td{\theta}}} \nc{\thetah}{{\hat{\theta}}} \nc{\vthetab}{{\bar{\vtht}}} \nc{\vthetat}{{\td{\vtht}}} \nc{\vthetah}{{\hat{\vtht}}}
\nc{\iotab}{{\bar{\iota}}}   \nc{\iotat}{{\td{\iota}}}   \nc{\iotah}{{\hat{\iota}}} \nc{\kappab}{{\bar{\kappa}}} \nc{\kappat}{{\td{\kappa}}}
\nc{\kappah}{{\hat{\kappa}}} \nc{\lmdb}{{\bar{\lmd}}}     \nc{\lmdt}{{\td{\lmd}}}     \nc{\lmdh}{{\hat{\lmd}}} \nc{\mub}{{\bar{\mu}}}
\nc{\mut}{{\td{\mu}}}       \nc{\muh}{{\hat{\mu}}} \nc{\nub}{{\bar{\nu}}}       \nc{\nut}{{\td{\nu}}}       \nc{\nuh}{{\hat{\nu}}}
\nc{\xib}{{\bar{\xi}}}       \nc{\xit}{{\td{\xi}}}       \nc{\xih}{{\hat{\xi}}} \nc{\pib}{{\bar{\pi}}}       \nc{\pit}{{\td{\pi}}}
\nc{\pih}{{\hat{\pi}}} \nc{\vpib}{{\bar{\vpi}}}     \nc{\vpit}{{\td{\vpi}}}     \nc{\vpih}{{\hat{\vpi}}} \nc{\rhob}{{\bar{\rho}}}
\nc{\rhot}{{\td{\rho}}}     \nc{\rhoh}{{\hat{\rho}}} \nc{\vrhob}{{\bar{\vrho}}}   \nc{\vrhot}{{\td{\vrho}}}   \nc{\vrhoh}{{\hat{\vrho}}}
\nc{\sigmab}{{\bar{\sigma}}} \nc{\sigmat}{{\td{\sigma}}} \nc{\sigmah}{{\hat{\sigma}}} \nc{\vsigmab}{{\bar{\vsgm}}} \nc{\vsigmat}{{\td{\vsgm}}}
\nc{\vsigmah}{{\hat{\vsgm}}} \nc{\taub}{{\bar{\tau}}}     \nc{\taut}{{\td{\tau}}}     \nc{\tauh}{{\hat{\tau}}} \nc{\upsilonb}{{\bar{\ups}}}
\nc{\upsilont}{{\td{\ups}}} \nc{\upsilonh}{{\hat{\ups}}} \nc{\phib}{{\bar{\phi}}}     \nc{\phit}{{\td{\phi}}}     \nc{\phih}{{\hat{\phi}}}
\nc{\vphib}{{\bar{\vphi}}}   \nc{\vphit}{{\td{\vphi}}}   \nc{\vphih}{{\hat{\vphi}}} \nc{\chib}{{\bar{\chi}}}     \nc{\chit}{{\td{\chi}}}
\nc{\chih}{{\hat{\chi}}} \nc{\psib}{{\bar{\psi}}}     \nc{\psit}{{\td{\psi}}}     \nc{\psih}{{\hat{\psi}}} \nc{\omegab}{{\bar{\omega}}}
\nc{\omegat}{{\td{\omega}}} \nc{\omegah}{{\hat{\omega}}}

\nc{\Gammab}{{\ovl{\Gamma}}}     \nc{\Gammat}{{\wtd{\Gamma}}}     \nc{\Gammah}{{\wht{\Gamma}}} \nc{\Deltab}{{\ovl{\Delta}}}
\nc{\Deltat}{{\wtd{\Delta}}}     \nc{\Deltah}{{\wht{\Delta}}} \nc{\Thetab}{{\ovl{\Theta}}}     \nc{\Thetat}{{\wtd{\Theta}}}
\nc{\Thetah}{{\wht{\Theta}}} \nc{\Lambdab}{{\ovl{\Lambda}}}   \nc{\Lambdat}{{\wtd{\Lambda}}}   \nc{\Lambdah}{{\wht{\Lambda}}}
\nc{\Xib}{{\ovl{\Xi}}}           \nc{\Xit}{{\wtd{\Xi}}}           \nc{\Xih}{{\wht{\Xi}}} \nc{\Pib}{{\ovl{\Pi}}}           \nc{\Pit}{{\wtd{\Pi}}}
\nc{\Pih}{{\wht{\Pi}}} \nc{\Sigmab}{{\ovl{\Sigma}}}     \nc{\Sigmat}{{\wtd{\Sigma}}}     \nc{\Sigmah}{{\wht{\Sigma}}}
\nc{\Upsilonb}{{\ovl{\Upsilon}}} \nc{\Upsilont}{{\wtd{\Upsilon}}} \nc{\Upsilonh}{{\wht{\Upsilon}}} \nc{\Phib}{{\ovl{\Phi}}}
\nc{\Phit}{{\wtd{\Phi}}}         \nc{\Phih}{{\wht{\Phi}}} \nc{\Psib}{{\ovl{\Psi}}}         \nc{\Psit}{{\wtd{\Psi}}}
\nc{\Psih}{{\wht{\Psi}}} \nc{\Omegab}{{\ovl{\Omega}}}     \nc{\Omegat}{{\wtd{\Omega}}}     \nc{\Omegah}{{\wht{\Omega}}}

\begin{document}

\maketitle

\begin{abstract}
We explore the nonperturbative aspects of the chiral algebras of $\CN = (0,2)$ sigma models, which perturbatively are intimately related to the
theory of chiral differential operators (CDOs).  The grading by charge and scaling dimension is anomalous if the first Chern class of the target
space is nonzero.  This has some nontrivial consequences for the chiral algebra.  As an example, we study the case where the target space is
$\CP^1$, and show that worldsheet instantons trivialize the chiral algebra entirely.  Consequently, supersymmetry is spontaneously broken in
this model.  We then turn to a closer look at the supersymmetry breaking from the viewpoint of Morse theory on loop space.  We find that
instantons interpolate between pairs of perturbative supersymmetric states with different fermionic numbers, hence lifting them out of the
supersymmetric spectrum.  Our results reveal that a ``quantum'' deformation of the geometry of the target space leads to a trivialization of the
kernels of certain twisted Dirac operators on $\CP^1$.
\end{abstract}

\thispagestyle{myheadings}
\let\tpg\thepage
\let\thepage\relax
\markboth{}{\normalsize\normalfont\hfill RUNHETC--2008--01}

\vfill

\pagebreak

\let\thepage\tpg

\section{Introduction}

Two-dimensional twisted $\CN = (0,2)$ sigma models were recently studied by Witten in~\cite{Witten:2005px}, where he showed that the
infinite-dimensional chiral algebra of local operators and the various perturbative characteristics of the models can be expressed in terms of
the sheaf of chiral differential operators (CDOs).  The theory of CDOs had first been introduced and developed in a series of seminal papers by
Malikov et al.~\cite{Malikov:1998dw, GMS}.  It has since found interesting applications in geometry and representation theory, such as mirror
symmetry~\cite{Bo} and the study of elliptic genera~\cite{BL, BL1, BL2}, just to name a few.  The mathematical relevance of twisted $(0,2)$
sigma models is thus clear in this respect.

Most existing discussions on the chiral algebras of $(0,2)$ sigma models have been confined within the realm of perturbation theory.  However,
one cannot ignore nonperturbative effects if one wishes to acquire a complete understanding of the full quantum theory.  Even though there have
been recent efforts \cite{Adams:2003zy} to study the sigma model beyond perturbation theory, the analysis was mainly confined to the finite
dimensional subsector of scaling dimension zero.  In this paper, we have taken a modest step towards a study of the nonperturbative aspects of
the \emph{full} chiral algebras of $(0,2)$ sigma models.  We analyze the model with target space $X = \CP^1$, where we find that worldsheet
instantons can change the picture radically: The perturbative chiral algebra of the model is completely trivialized by instantons.
Consequently, supersymmetry is spontaneously broken in this model.  This nonperturbative phenomenon can also be understood from the perspective
of Morse theory on loop space, where it is shown that instantons lift pairs of perturbative states from the supersymmetric spectrum.  Moreover,
our results suggest that under a ``quantum'' deformation of the geometry of the target space, the kernels of certain twisted Dirac operators on
$\CP^1$ will become empty.

The structure of the paper is as follows.  In Section 2, we will discuss twisted $\CN = (0,2)$ nonlinear sigma models in two dimensions.  After
briefly reviewing the general features of these models, we will introduce their chiral algebras and the cohomology of physical states.  These
will be the main objects of interest in this paper.  We will also review the anomalies of the models, which will be relevant to the discussions
in the later sections.

In Section 3, we will elaborate on the perturbative chiral algebras and their relation to the sheaves of CDOs.  We will start with the chiral
algebras in the classical theory, where one has the \v Cech--Dolbeault isomorphism to formulate them as the \v Cech cohomology of a certain
sheaf over the target space.  We will then consider perturbative quantum corrections.  As we will see, the chiral algebras can still be
expressed in terms of \v Cech cohomology.  We will explain how one can carry out the actual computations by ``gluing'' free theories over the
target space.

In Section 4, the chiral algebra of the $\CP^1$ model will be studied in the full quantum theory.  To this end, we will first determine the
perturbative chiral algebra.  This will also serve as an illustration of the general concepts developed in the previous section.  We will then
extend our analysis beyond perturbation theory, taking into account the presence of worldsheet instantons.  By analyzing the operator product
expansion between the supercurrent and a certain local operator in an instanton background, we find that the perturbative chiral algebra is
completely trivialized by instanton effects.  This in particular implies the spontaneous breaking of supersymmetry.

In Section 5, we will turn to a closer look at the supersymmetry breaking from the viewpoint of Morse theory on loop space.  There we will see
that the Fock space structure is altered by a spectral flow which arises when the closed string wraps the $\CP^1$ target as it propagates.  This
allows instantons to interpolate between pairs of perturbative supersymmetric states with different fermionic numbers, hence lifting them out of
the supersymmetric spectrum.

\section{Chiral Algebras of $(0,2)$ Models}
\label{chiral-algebra}

In this section, we review $\CN = (0,2)$ nonlinear sigma models in two dimensions and their chiral algebras.  We will restrict ourselves to the
simplest case, in which there are no left-moving fermions coupled to the gauge bundle over the target space.

\subsection{The Model}
\label{model}

Let the worldsheet $\Sigma$ be a Riemann surface, and the target $X$ a compact hermitian manifold.  We will formulate the theory in the twisted
case, so that there exists a global supersymmetry for an arbitrary choice of $\Sigma$.  The bosonic field $\phi$ is a map from $\Sigma$ to $X$,
while the right-moving fermions $\alpha$, $\rho$ are sections
\begin{equation}
\alpha^\ib \in \Gamma(\phi^*\overline{TX}), \qquad \rho_\zb^i \in \Gamma(\Kb_\Sigma \otimes \phi^*TX),
\end{equation}
where $K_\Sigma$ is the canonical line bundle of $\Sigma$.  Under the right-moving $R$ symmetry, $\alpha$ has charge $1$ and $\rho$ has charge
$-1$.

There are several requirements which $\Sigma$ and $X$ must meet.  First, the model suffers from the sigma model anomalies unless certain
conditions are satisfied~\cite{Moore:1984ws}.  In addition to the familiar one which leads to the condition $p_1(X)/2 = 0$, the twisting
introduces a further anomaly if $c_1(X)$ and $c_1(\Sigma)$ are both nonzero \cite{Witten:2005px}.  If one wishes to study a theory on a target
space with $c_1(X) \neq 0$, such as $X = \CP^1$, one must consider a worldsheet with trivial canonical line bundle.  Second, we assume $X$ to be
spin, or equivalently, $c_1(X) \equiv 0$ (mod 2).  From the canonical quantization viewpoint, this condition is interpreted as the orientability
of the loop space $LX = \Map(S^1,X)$.  Together with $p_1(X)/2 = 0$, this implies the existence of the Dirac operator on $LX$
\cite{Witten:1986bf, Witten:1987cg}.  Finally, $\CN = (0,2)$ supersymmetry requires the associated $(1,1)$-form $\omega$ of $X$ to obey the
equation $\del\delb\omega = 0$ \cite{Witten:2005px}.  Such a real $(1,1)$-form can be written locally as $\omega = i(\del\Kb - \delb K)$  for
some $(1,0)$-form $K$.  The target metric is then given by $g_{i\jb} = \del_i\Kb_\jb + \del_\ib K_j$.

After twisting, the supercharge $Q_+$ becomes a $(0,1)$-form on $\Sigma$, while $\Qb_+$ becomes a worldsheet scalar.  Associated with the latter
is the fermionic symmetry generated by~$\epsilon\delta$, where $\epsilon$ is a scalar grassmannian, and $\delta$ transforms the fields by
\begin{equation}
\label{Q}
\begin{alignedat}{3}
\delta\phi^i     &=  0, &\qquad
\delta\phi^\ib   &= \alpha^\ib, \\
\delta\rho_\zb^i &= -\del_\zb\phi^i, & \delta\alpha^\ib &=  0.
\end{alignedat}
\end{equation}
As one can see from the above transformation law, the fermionic symmetry is nilpotent, i.e., $\delta^2 = 0$, and increases the $R$ charge by
one.

An action of the form $\int_\Sigma \delta V$ for a $(1,1)$-form $V$ on $\Sigma$ of charge $-1$ is manifestly invariant under the fermionic and
$R$ symmetries.  A choice compatible with $\CN = (0,2)$ supersymmetry is the sum of the following two terms:
\begin{equation}
S_0 = \int_\Sigma d^2z \,
      \delta\bigl(-g_{i\jb} \rho_\zb^i \del_z\phi^\jb\bigr), \qquad
S_T = \int_\Sigma d^2z \, \delta\bigl(-T_{ij}\rho_\zb^i\del_z\phi^j\bigr).
\end{equation}
Here, $d^2z = idz \wedge d\zb$ and $T = -\del K$ is a $(2,0)$-form defined locally on $X$.  This leads to the action
\begin{equation}
\label{S} S_0 + S_T = \int_\Sigma d^2z \, g_{i\jb}
  \bigl(\del_\zb\phi^i \del_z\phi^\jb + \rho_\zb^i D_z\alpha^\jb \bigr)
  -i\int_\Sigma \phi^*T
\end{equation}
with the twisted covariant derivative
\begin{equation}
D_z\alpha^\jb = \del_z\alpha^\jb
  + \del_z\phi^\ib g^{\jb l} \del_\kb g_{l\ib} \alpha^\kb
  + \del_z\phi^i g^{\jb l} \del_\kb T_{li} \alpha^\kb.
\end{equation}
In general, $T$ is not well defined as a $(2,0)$-form.  Rather, it is a two-form gauge field, whose gauge transformation is given by $T \to T +
\tau$ for a closed local $(2,0)$-form $\tau$.

One may also add to the action a topological term constructed as follows.  Let $B$ be a closed two-form on $X$.  Then, the functional
\begin{equation}
S_B = \int_\Sigma \phi^* B
\end{equation}
is topological, in the sense that it depends only on the cohomology class $[B]$ and the homology class $[\phi(\Sigma)]$.  As such, it is
invariant under any continuous transformations, and in particular, invariant under the fermionic and $R$ symmetries.  The topological action
$S_B$ vanishes in perturbation theory, where one considers only homotopically trivial maps.  Nonperturbatively, it serves as the weights
assigned to the instanton sectors.  In the case of a K\"ahler target space, one typically takes the real part of $B$ to be the K\"ahler form.

\subsection{Chiral Algebra and Cohomology of States}

The generator $Q = \Qb_+$ of the fermionic symmetry acts on local operators by the commutator: $\{Q,\CO\} = \delta\CO$.  From the nilpotency of
the fermionic symmetry together with the fact that $Q$ has charge $1$, we can immediately construct the $Q$-cohomology of local operators graded
by $R$ charge.

While the cohomology of local operators was an automatic consequence of the nilpotency of the fermionic symmetry, the cohomology of
\emph{states} is not.  The nilpotency only implies that $Q^2$ is a central charge, i.e., it commutes with any operator in the theory.  In our
case, however, the central charge is absent---this is simply because given a generic hermitian metric, there is no candidate for such a
conserved quantity.  The fermionic operator $Q$ is nilpotent and a worldsheet scalar, and hence, it is sometimes referred to as a BRST operator.

The relation between the cohomology of local operators and states is that the latter is a module over the former:
\begin{equation}
[\CO] \cdot [\ket{\Psi}] = [\CO\ket{\Psi}].
\end{equation}
In particular, the cohomology of states is trivial if that of local operators is so: In such a case, the constant operator $1$ is $Q$-exact,
thereby an operator $W$ exists such that $1 = \{Q,W\}$.  Consequently, for any state $\ket{\Psi}$ with $Q\ket{\Psi} = 0$, one has $\ket{\Psi} =
Q(W\ket{\Psi})$.  As we will see later, this scenario is indeed realized for $X = \CP^1$.

Let us take a closer look at the cohomology of local operators.  Inside the correlation functions, local operators carry their insertion point
on the worldsheet.  We will indicate this by writing them as $\CO(z,\zb)$.  As shown in Section~\ref{CDO}, perturbatively the generator
$\Lb_{-1} = \del_\zb$ of antiholomorphic translations is trivial in the cohomology, i.e., $\Lb_{-1} = \{Q,W\}$ for some~$W$.  Perturbative
cohomology classes thus vary holomorphically on $\Sigma$, since if a local operator $\CO$ represents a cohomology class, we have $\del_\zb \CO
\propto \{Q,\{W,\CO\}\}$ by the Jacobi identity.  Furthermore, two cohomology classes can be multiplied by the operator product expansion (OPE),
and the resulting operator again lies in the cohomology:
\begin{equation}
[\CO_a(z_a)] \cdot [\CO_b(z_b)] \sim \sum_c f^c{}_{ab}(z_a - z_b) [\CO_c(z_b)].
\end{equation}
This structure---the holomorphic $Q$-cohomology with OPE---is called the \emph{chiral algebra} of the model, which we will henceforth denote by
$\CA$.  As we will see, this is intimately related to the chiral algebra in the sense of conformal field theories in two dimensions, i.e., the
algebra of holomorphic symmetry currents.

The chiral algebra forms the ``quasi-topological'' sector of the theory, which will be called the BRST sector in the present paper.  The
correlation functions of $Q$-closed operators,
\begin{equation}
\label{BRST} \Bigvev{\prod_a \CO_a(z_a,\zb_a)}; \quad \{Q,\CO_a\} = 0,
\end{equation}
depend only on the cohomology classes $[\CO_a]$: If one of them is exact, then it becomes the one point function of a $Q$-exact operator, i.e.,
$\vev{\{Q,W\}}$, which vanishes by virtue of the fermionic symmetry.  This in particular means that the correlation function~\eqref{BRST} is a
holomorphic function of the insertion points.  Furthermore, it is invariant under a smooth deformation of the hermitian structure of $X$, since
such a perturbation just brings down a $Q$-exact term.  A similar argument also shows the invariance of the correlation functions under a
deformation of the two-form field $T$ by a \emph{globally} defined $(2,0)$-form.

\subsection{Anomalies}
\label{anomalies}

We will now discuss the anomalies of our model.  For this purpose, it will be helpful to review how the path integral measure is constructed,
and how correlation functions are defined within this context.  This also prepares us for the subject of the next section, the sheaf of chiral
differential operators.

\subsubsection{Localization}
\label{localization}

The localization principle \cite{Witten:1991zz} states that path integrals in the BRST sector (or in the large volume limit) localize around
instanton configurations, which in our case are holomorphic maps from~$\Sigma$ to~$X$.  Let~$\CC = \Map(\Sigma, X)$ be the bosonic configuration
space, and $\CM \subset \CC$ be the instanton moduli space, i.e., the space of holomorphic maps from~$\Sigma$ to~$X$.  Then, we can reduce the
integration region to an arbitrary neighborhood~$\CM' \supset \CM$.  This provides a considerable simplification of computations.

The fermions~$\rho$ and~$\alpha$ are respectively sections of~$\Kb_\Sigma \otimes \phi^*\TXb$ and~$\phi^*\TXb$, hence in order to define a
fermionic path integral measure, we must introduce local frames on these bundles.  To this end, let us introduce unitary eigenmodes of the
laplacians:
\begin{equation}
\begin{alignedat}{2}
\Delta_F(\phi) u_{0,r}(z,\zb; \phi) &= 0, &\qquad
\Delta_F(\phi) u_n(z,\zb; \phi) &= \lambda_n(\phi) u_n(z,\zb; \phi), \\
\Delta_F^\dagger(\phi) v_{0,s}(z,\zb; \phi) &= 0, &\qquad
\Delta_F^\dagger(\phi) v_n(z,\zb; \phi) &= \lambda_n(\phi) v_n(z,\zb; \phi), \\
\end{alignedat}
\end{equation}
where $\Delta_F = \Db^\dagger\Db$ with $\Db = d\zb D_\zb$, and the nonzero modes are paired by the relation $v_n = \lambda_n^{-1/2} \Db u_n$.  On a sufficiently small open set $U \subset \CM'$, we can take the eigenmodes to vary smoothly in $\phi$.%
\footnote{We are assuming here that there is no singularity in the instanton moduli space.  In particular, the dimension of $\CM$ does not jump
within a connected component.}
One can then expand the fermions as
\begin{equation}
\begin{split}
\rho(z,\zb; \phi)
&= \sum_s b_0^s v_{0,s}(z,\zb; \phi) + \sum_n b^n v_n(z,\zb; \phi), \\
\alpha(z,\zb; \phi) &= \sum_r c_0^r \ub_{0,r}(z,\zb; \phi) + \sum_n c^n \ub_n(z,\zb; \phi),
\end{split}
\end{equation}
where $b_0^s$, $c_0^r$, $b^n$, $c^n$ are grassmannian variables.  The fermionic path integral measure is then defined, locally on~$U$, by the
formal product
\begin{equation}
\label{FPM} \prod_{r,s,n} db_0^s dc_0^r db^n dc^n.
\end{equation}
Notice that the eigenmodes in terms of which the fermions are expanded are themselves functions of fluctuating quantum fields.  This point will
be important in Section~\ref{CP1}.

To construct a bosonic path integral measure, let us choose~$\CM'$ to be a tubular neighborhood of~$\CM$, so that it is diffeomorphic to the
normal bundle~$N\CM$.  A point $\phi \in \CM'$ can then be identified, via the exponential map, with a vector in~$N_{\phi_0}\CM$ at the nearest
point $\phi_0 \in \CM$.  This vector can be identified with a vector field pointing in the directions normal to~$\CM$, hence, it represents a
fluctuation around an instanton.  We denote the~$(1,0)$ and~$(0,1)$ components of this vector field by~$\vphi(z,\zb; \phi_0) \in \phi_0^*TX$
and~$\vphib(z,\zb; \phi_0) \in \phi_0^*\TXb$, respectively.  Note that $\phi_0^*TX$ is a holomorphic vector bundle over~$\Sigma$, and thus the
Dolbeault operator $\delb$ acts naturally on the space of its sections.  The eigenmodes~$u_{0,r}$ and~$u_n$ reduce on~$\CM$ to eigenmodes of the
laplacian~$\delb^*\delb$.  The zero modes satisfy $\delb u_{0,r} = 0$, and therefore, they are holomorphic and tangent to~$\CM$.  The normal
directions are spanned by the nonzero modes:
\begin{equation}
\vphi(z,\zb; \phi_0) = \sum_n a^n u_n(z,\zb;\phi_0), \qquad \vphib(z,\zb; \phi_0) = \sum_n \ab^n \ub_n(z,\zb;\phi_0).
\end{equation}
The bosonic path integral measure on~$U \cap \CM$ is then
\begin{equation}
\label{BPM} d\CM \prod_{n} da^n d\ab^n.
\end{equation}
Here, $d\CM$ is the canonical volume form of $\CM$.  The total path integral measure is the product of the bosonic and fermionic measures.  Path
integral measures defined locally this way can then be glued patch by patch over all of~$\CM$ without any obstruction, if the sigma model
anomalies are absent \cite{Moore:1984ws}.

For the path integral measure to be well defined, we must regularize the formal expressions~\eqref{FPM} and~\eqref{BPM}.  We will do this by
imposing an ultraviolet cutoff~$\Lambda_{\text{UV}}$, so that only the modes with $\lambda_n < \Lambda_\text{UV}$ will be integrated.  The
regularized measure is invariant under the fermionic symmetry, as one can verify explicitly by computing the Berezinian.  The nilpotency of~$Q$
also follows as mentioned in Section~\ref{model}.  The other ingredients required to define the cohomology of local operators and states are the
$R$~charge and the scaling dimension.  We now turn to a discussion of their anomalies.

\subsubsection{Charge and Scale Anomalies}
\label{charge-anomalies}

In general, the configuration space~$\CC$ splits into connected components as $\CC = \bigcup_\beta \CC_\beta$, which are labeled by the homology
class $\beta = [\phi(\Sigma)]$ of the embedding of the worldsheet into the target space.  On each component, the above construction is applied
to define the path integral measure, and once this is done, the correlation functions take the form of the sum over the instanton sectors:
\begin{equation}
\bigvev{\cdots} = \sum_\beta e^{-t_\beta} \bigvev{\cdots}_\beta.
\end{equation}
In this expression, $e^{-t_\beta}$ is the weight assigned to~$\CC_\beta$ given by the topological action~$S_B$, and $\vev{\dots}_\beta$ is the
correlation function computed in that sector.

As a result of the paring of nonzero modes, the nonzero mode part of the fermionic path integral measure \eqref{FPM} is neutral under the $R$ symmetry.  On the other hand, the zero mode part has charge equal to the number of $\rho$ zero modes $v_{0,s}$ minus the number of $\alpha$ zero modes $\ub_{0,r}$, i.e., minus of the index of $\delb$ twisted by $\phi^*TX$.  On a compact Riemann surface $\Sigma$ with genus $g$, the index for the $\beta$-instanton sector is given by%
\footnote{The first term is absent in the physical (i.e., untwisted) models, where the fermions are worldsheet spinors.}
\begin{equation}
\label{index} (1-g) \cdot \dim_\C X + \ip{\beta}{c_1(X)}.
\end{equation}
Thus, the $R$ symmetry is anomalous.  Due to this charge anomaly, the correlation function $\vev{\cdots}_\beta$ vanishes unless the total charge
of the inserted operators is equal to~\eqref{index}.  Furthermore, if $c_1(X) \neq 0$, the continuous $U(1)_R$ symmetry is broken to the
discrete $\Z_{2k}$ symmetry, where $2k$ is the greatest common divisor of $\ip{\beta}{c_1(X)}$.  (Recall the condition $c_1(X) \equiv 0$ (mod
$2$).)  As a result, it is possible for two operators $\CO_1$ and~$\CO_2$, whose charges differ by a multiple of $2k$, to define the same
cohomology class in the \emph{full} correlation function.  Thus, the grading of the chiral algebra by charge reduces to a $\Z_{2k}$ grading in
the presence of instantons.

By the same token, scale invariance is also anomalously broken by instantons.  Under a scaling of the metric $h_{ab} \to \Lambda^2 h_{ab}$, the
eigenmodes $u_{0,r}$, $u_n$ must scale as $u \to \Lambda u$ to maintain unitarity.  This induces the transformation $da \to \Lambda^{-1} da$, \
$dc \to \Lambda dc$.  The zero mode part $\CD\mu_0$ of the path integral measure (which includes $d\CM$) then scales as
\begin{equation}
\CD\mu_0|_{\CM'_\beta} \to \Lambda^{-\dim_\C\CM_\beta} \CD\mu_0|_{\CM'_\beta}.
\end{equation}
The scaling of the nonzero modes needs to be regularized.  In the case of conformally invariant theories, this will lead to the familiar
conformal anomaly proportional to the scalar curvature of the worldsheet.  If one only works in one of the instanton sectors, the constant
factor which arises from a scaling of the path integral measure can be ignored by considering normalized correlation functions.  However, the
presence of the scale anomaly indicates that the grading by dimension will be violated by instantons.  This has some nontrivial consequences for
the cohomology in a nonperturbative setting, as we will see in Section~\ref{CP1}.

\section{Perturbative Chiral Algebra and CDO}
\label{CDO}

Our discussion so far has been concerned with the general features of chiral algebras, which apply to any choice of target space.  In this
section, we focus on the relation between the chiral algebra and the geometry of the target space.  We will start with the classical chiral
algebra, and then proceed to discuss its quantum counterpart in perturbation theory.  The main goal of this section is to understand how the
perturbative chiral algebra can be computed from a certain sheaf derived from free theories, namely, the sheaf of chiral differential operators.
We refer the reader to \cite{Witten:2005px} for more discussions.  For a generalization where left-moving fermions are present, see
\cite{Tan:2006qt}.

\subsection{Classical Chiral Algebra}

Classically, our model is conformally invariant.   To compute the stress tensor, we first couple the theory to the worldsheet metric $h_{ab}$ as
\begin{equation}
S = \int_\Sigma \sqrt{h} d^2x h^{ab}
    \delta\bigl(g_{i\jb} \rho_a^i \del_b\phi^\jb
                + T_{ij}\rho_a^i\del_b\phi^j\bigr),
\end{equation}
and take the variation $\delta h_{ab}$.  We find
\begin{equation}
\begin{aligned}
T_{zz}     &= g_{i\jb}\del_z\phi^i \del_z\phi^\jb
              + T_{ij}\del_z\phi^i \del_z\phi^j, \\
T_{\zb\zb} &= \bigl\{Q, g_{i\jb} \rho_\zb^i \del_\zb\phi^\jb
                        + T_{ij}\rho_\zb^i\del_z\phi^j\bigr\},
\end{aligned}
\end{equation}
and $T_{z\zb} = T_{\zb z} = 0$.  The components of the stress tensor commute with $Q$, thus they descend to well-defined operators in the
cohomology.  This is manifestly true for $T_{\zb\zb}$ since it is expressed as a $Q$-exact operator.  As for $T_{zz}$, it is indeed $Q$-closed
upon using the equation of motion $D_z\alpha^\ib = 0$.

Since $T_{\zb\zb}$ vanishes in the chiral algebra, so do the generators $\Lb_n$ of the antiholomorphic conformal transformations $\zb \to \zb +
\epsilon \zb^{n+1}$.  The special case of this is $\Lb_{-1} = \del_\zb$, as we have already discussed in Section~\ref{chiral-algebra}.  On the
other hand, the condition from $\Lb_0$ leads to a considerable simplification of the chiral algebra: If a local operator $\CO$ has a nonzero
right-moving dimension, $\{\Lb_0,\CO\} = \hb\CO$ with $\hb \neq 0$, then it is $Q$-exact.  Thus, the chiral algebra consists only of local
operators with $\Lb_0 = 0$.  This means that $\rho$ and $\zb$-derivative of any fields do not enter the chiral algebra, as they have a positive
$\Lb_0$.  We will hereafter refer to the eigenvalue of $L_0$ simply as the dimension of the operator.

Thus, local operators relevant to the classical chiral algebra are, locally on $\Sigma$ and $X$, functions of the form
\begin{equation}
\label{CO} \CO(\phi, \del_z\phi, \del_z^2\phi, \dotsc; \phib, \del_z\phib, \del_z^2\phib, \dotsc; \alpha).
\end{equation}
Here we have used the equation of motion to replace $\del_z\alpha$ with other variables.  By locality, the expansion of $\eqref{CO}$ in
$\del_z^n\phi$ and $\del_z^n\phib$ must stop at a finite order.  The expansion in $\alpha$ stops at the order $d = \dim_\C X$ by Fermi
statistics, as there are $d$ zero modes for $\alpha$ in the zeroth instanton sector $\CM_0 \iso X$, i.e., the space of constant maps from
$\Sigma$ to $X$.  However, the dependence on the target coordinates $(\phi, \phib)$ is not constrained.  Therefore, a local operator can take
the general form
\begin{equation}
\label{LO} \CO(\phi,\phib)_{j\dotsm;k\dotsm;\dotsm}{}^{l\dotsm;m\dotsm;\dotsm}{}_{\ib_1\dotsm\ib_q} \del_z\phi^j \dotsm \del_z^2\phi^k \dotsm
\del_z\phib_l \dotsm \del_z^2\phib_m \dotsm \alpha^{\ib_1} \dotsm \alpha^{\ib_q}.
\end{equation}
Note that we have used the hermitian metric to lower the indices.

Globally over $X$, but locally on $\Sigma$, the operator \eqref{LO} can be interpreted as a $(0,q)$-form taking values in a holomorphic vector
bundle over~$X$ constructed from tensor products of $TX$ and $T^*X$.  As such, we can take the space of local operators to be a vector bundle $F
= V \otimes \wedge^\bullet\overline{T^*X}$, where $V$ is a hermitian vector bundle given by the direct sum of possible vector bundles that
appear as a coefficient in the expression~\eqref{LO}.  Globally on both $\Sigma$ and $X$, the section \eqref{LO} of $F$ must glue consistently
over $\Sigma$.  Classically, this is always possible by using worldsheet covariant derivatives.

The bundle $F$ is graded by charge and dimension.  We denote the subbundle of charge $q$ and dimension $h$ by $F^{q,h}$.  The chiral algebra inherits the grading by charge and dimension, so that one can write $\CA = \bigoplus_{q,h \geq 0} \CA^{q,h}$.  Let us assume $T_{ij} = 0$ for simplicity; such is the case for sigma models on K\"ahler manifolds with $\CN = (0,2)$ supersymmetry.  From the transformation law \eqref{Q}, the fact that $\rho$ is absent from the bundle, and the equation of motion $D_z\alpha^\ib = 0$, we see that $Q$ acts on $F^{\bullet,h}$ as the Dolbeault operator~$\delb$.%
\footnote{If the two-form field $T$ is present, the equation of motion for $\alpha$ is modified and the action of $Q$ is twisted by the
$(2,1)$-part of the field strength $\CH = dT$.}
Thus, the $q$th $Q$-cohomology group of dimension $h$ is, classically, given by the Dolbeault cohomology group $H_\delb^q(X, F^{\bullet,h})$.
Since there are no short distance singularities at the classical level, the multiplication structure of the chiral algebra is given by the naive
product of operators, which coincides with the ring structure of the Dolbeault cohomology given by the wedge product.

A powerful means of computing Dolbeault cohomology is provided by \v Cech cohomology.  Let $\CF^{q,h}$ be the sheaf of local sections of
$F^{q,h}$, and $\CZ^{q,h}$ be the subsheaf of $\CF^{q,h}$ of $\delb$-closed sections.  The \v Cech--Dolbeault isomorphism states that
\begin{equation}
H_\delb^q(X, F^{\bullet,h}) \iso \check H^q(X,\CZ^{0,h}).  \label{CD}
\end{equation}
It will be useful for our upcoming discussion to briefly recall how this isomorphism is realized.  The first step is the $\delb$-Poincar\'e
lemma, by which one obtains the short exact sequence of sheaves:
\begin{equation}
0 \longto \CZ^{q-1,h}
  \longto \CF^{q-1,h}
  \stackrel{\delb}{\longto} \CZ^{q,h}
  \longto 0.
\end{equation}
This in turn induces the long exact sequence of cohomology:
\begin{equation}
\dotsb \to \check H^p(\CZ^{q-1,h}) \to \check H^p(\CF^{q-1,h}) \stackrel{\delb}{\to} \check H^p(\CZ^{q,h}) \to \check H^{p+1}(\CZ^{q-1,h}) \to
\dotsb.
\end{equation}
Now, since $\CF^{q,h}$ is a fine sheaf, $\check H^p(X, \CF^{q,h}) = 0$ for $p > 0$.  The long exact sequence then produces a chain of
isomorphisms:
\begin{equation}
\label{COI} \check H^q(\CZ^{0,h}) \iso \check H^{q-1}(\CZ^{1,h}) \iso \dotsb \iso \check H^1(\CZ^{q-1,h}) \iso \check H^0(\CZ^{q,h})/\delb
\check H^0(\CF^{q-1,h}).
\end{equation}
This establishes the \v Cech--Dolbeault isomorphism $\check H^q(X,\CZ^{0,h}) \iso H_\delb^q(X, F^{\bullet,h})$.

\subsubsection{Example: $c_1(X) \neq 0$}
\label{c1}

We will now illustrate the use of the \v Cech--Dolbeault isomorphism.  To this end, let us consider, as a relevant example, the case when the
target space has nonzero first Chern class.  We now wish to ascertain if there is anything in the chiral algebra that can be associated with
$c_1(X)$.

Let $\set{U_a}$ be a good cover, i.e., all nonempty finite intersections of its open sets are diffeomorphic to $\R^n$.  We choose holomorphic
transition functions $\set{f_{ab}}$ of the canonical line bundle $K_X$ on overlaps $\set{U_a \cap U_b}$.  On a triple overlap $U_a \cap U_b \cap
U_c$, the transition functions satisfy the cocycle condition:
\begin{equation}
\label{cocycle} \delta f_{abc} = f_{bc} f_{ca} f_{ab} = 1.
\end{equation}
Hence, $\set{f_{ab}}$ defines a cohomology class in $\check H^1(X,\CO^\times)$.  This is mapped to an integral cohomology class $c_1(X)$ under
the homomorphism $H^1(X,\CO^\times) \to H^2(X,\Z)$ induced by the short exact sequence
\begin{equation}
0 \longto \Z \longto \CO \longto \CO^\times \longto 1.
\end{equation}
The map $\CO \to \CO^\times$ is given by $f \mapsto \exp(2\pi if)$.

To discuss the chiral algebra, the cocycle condition \eqref{cocycle} needs to be written additively.  One may try to do this by the map $f_{ab}
\mapsto \tilde f_{ab} = \log f_{ab}/2\pi i$, which is well defined since we are using a good cover.  Naively, the right-hand side seems to be
mapped to zero and $\set{\tilde f_{ab}}$ defines a cocycle; however, because of the nontrivial topology of the manifold, one only has
\begin{equation}
\label{c1-cocycle} \delta\tilde f_{abc} = \tilde f_{bc} + \tilde f_{ca} + \tilde  f_{ab} \in \Z.
\end{equation}
In fact, $\set{\delta\tilde f_{abc}}$ gives an explicit representation of $c_1(X) \in H^2(X,\Z)$.

To obtain a cocycle associated to $\set{f_{ab}}$, we make use of the additional degree of freedom peculiar to the field theory.  We take the
derivative of \eqref{c1-cocycle}, and set
\begin{equation}
\theta_{ab} = \del_z\tilde f_{ab}
            = \frac{1}{2\pi i} f_{ab}^{-1} \del_z f_{ab}.
\end{equation}
Then, $\set{\theta_{ab}}$ defines a holomorphic cocycle and therefore a cohomology class $[\set{\theta_{ab}}] \in \check H^1(X,\CZ^{0,1})$,
which in turn is mapped via the \v Cech--Dolbeault isomorphism to a cohomology class $[\theta] \in H_\delb^1(X,F^{\bullet,1}) \iso \CA^{1,1}$.
Applying $\del_z$ repeatedly, one obtains cohomology classes of higher dimensions, $[\del_z\theta]$, $[\del_z^2\theta]$, etc.  These are the
elements of the chiral algebra that are associated with $c_1(X)$.

Let us find an explicit representation of $[\theta]$ in $\CA^{1,1}$.  First, we note that $\theta_{ab}$ can be written as $\theta_{ab} =  W_b -
W_a$, where
\begin{equation}
W = -\frac{1}{2\pi i} \del_z\phi^i \del_i \log G,
\end{equation}
and $G = \det(g_{i\jb})$ is the determinant of the hermitian metric.  This does not mean that $[\theta_{ab}]$ is trivial in the \v Cech
cohomology, since $W_a$ are not holomorphic.  Rather, $W_a$ transform by holomorphic transition functions, so that $\delb W_a = \delb W_b$.
Hence, $\set{\delb W_a}$ defines a cohomology class in $\check H^0(X, \CZ^{1,h})/\delb \check H^0(X, \CF^{0,h}) \iso
H_\delb^1(X,F^{\bullet,1})$, which gives the desired representation of $\theta$.  Using the fact that the Ricci form $\CR = iR_{i\jb} d\phi^i
\wedge d\phi^\jb$ of a hermitian manifold can be computed as $\CR = -i\del\delb\log G$, we find that
\begin{equation}
\theta = \delb W_a
       = \frac{1}{2\pi i} R_{i\jb} \del_z\phi^i \alpha^\jb.
\end{equation}
We see from this expression that $\theta$ indeed corresponds to a cohomology class in $\CA^{1,1}$.

\subsection{Perturbative Chiral Algebra}

\label{PCA}

We now take up the issue of perturbative quantum corrections to the chiral algebra.  After reviewing the general structure of perturbative
corrections, we will elaborate on the sheaves of CDOs, which allow one to compute perturbative chiral algebras in the language of \v Cech
cohomology.

\subsubsection{Structure of Perturbative Chiral Algebras}

As we have seen in Section~\ref{anomalies}, the chiral algebra is still defined after quantum corrections are taken into account.
Perturbatively, we work only in $\CM_0 \iso X$, so we still have the $\Z$ grading given by $R$ charge.  We need further arguments to justify the
grading by dimension.  To this end, let us first look at the scale invariance of the cohomology.

Although the path integral measure scales by an overall factor under a global Weyl transformation, this can be neglected if we consider the
normalized correlation functions.  The actual problem is the change in the integration region in the field space, arising from the fact that the
ultraviolet cutoff $\Lambda_{\text{UV}}$ is fixed while the eigenvalues of the laplacians scale.  However, the integration region is restored by
scaling $\Lambda_{\text{UV}}$ appropriately.  Such sliding of $\Lambda_{\text{UV}}$ is equivalent to integrating out the high-frequency modes.
With appropriate field redefinitions, this will result in the renormalization of the metric \cite{Friedan:1980jm, AlvarezGaume:1981hn}.  Both of
these operations can be done within the $Q$-commutator in the action.  Thus, the correlation functions in the BRST sector is scale invariant in
perturbation theory.  As in the most examples in two dimensions \cite{Polchinski:1987dy}, our theory is also conformally invariant.  In fact,
$T_{z\zb}$ is $Q$-exact.

The fact that $T_{z\zb}$ is $Q$-exact implies that there exist operators $L_n$ and $\Lb_n$, which when restricted to their action on $Q$-closed
operators, generate conformal transformations.  It does \emph{not} mean, however, that conformal transformations leave $Q$-closed operators in
the cohomology.  Indeed, $T_{zz}$ was $Q$-closed on shell classically, but it no longer lies in the cohomology since perturbative corrections
modify the equations of motion.  Thus, $L_n$ in general destroy cohomology classes in perturbation theory.  On the other hand, $T_{\zb\zb}$
remains $Q$-exact in perturbation theory, so the charges of $\Lb_n$ vanish in the cohomology.  It follows that we can still use the same bundle
$F$ of local operators in the perturbative quantum theory as we did in the classical theory.

We can now show that the grading by dimension is preserved perturbatively.  Since $L_0 - \Lb_0$ generates Euclidean rotations on the worldsheet,
it can take only integer values.  We know that $\Lb_0 = 0$ in the cohomology, hence $L_0$ must be an integer that is protected from perturbative
corrections.  We have therefore justified the grading by dimension of the perturbative chiral algebra.

In fact, $L_{-1} = \del_z$ is also well defined in the cohomology.  To see this, consider first the flat worldsheet $\Sigma = \C$.  In this
case, the theory admits a global translation $z \to z + a$, $\zb \to \zb + \ab$.  This is a global symmetry which commutes with the fermionic
symmetry.  Since $[\Lb_{-1}, Q] = 0$, it follows that $[L_{-1}, Q] = 0$.  This statement remains true for a general Riemann surface, since there
are no possible corrections by tensors of correct type that survive in the continuum limit $\Lambda_{\text{UV}} \to \infty$.  This in turn
constrains the possible perturbative corrections to the commutator $[Q, T_{zz}]$.  If we plug into the equation $[Q,L_{-1}] = 0$ the expression
of $L_{-1}$, then the term involving $T_{z\zb}$ vanishes, and we are left with
\begin{equation}
\oint \! dz \, [Q, T_{zz}] = 0.
\end{equation}
This implies that $[Q, T_{zz}] = \del_z W$ for some $W$.  The conservation of $R$ charge, left and right-moving dimensions, and the power
counting on the target geometry leave only two possibilities at one-loop: $W \propto R_{i\jb} \del_z\phi^i\alpha^\jb$ or $g_{i\jb}R
\del_z\phi^i\alpha^\jb$.  We will see that the answer is the former, much as the one-loop correction to the metric is proportional to the Ricci
curvature.

\subsubsection{The \v Cech--$Q$ Isomorphism}

Our next task is to furnish a sheaf-theoretic interpretation of the perturbative chiral algebra that we have been discussing thus far.
Classically, the sheaf-theoretic interpretation was established via the \v Cech--Dolbeault isomorphism \eqref{CD}.  In the quantum theory, this
must be replaced by the \v Cech--$Q$ isomorphism:
\begin{equation}
\label{cech-q} H_Q^q(X,F^{\bullet,h}) \iso \check H^q(X,\CZ^{0,h}).
\end{equation}
In order for \eqref{cech-q} to hold, two points must be addressed: First, we need to ensure that $Q$ induces a well-defined morphism on the
sheaf of local operators.  Second, we must show that the argument that led us to the classical isomorphism is still valid.

The first point is easy.  Although the action of $Q$ gets corrected perturbatively, it can be restricted to an arbitrary small open set in $X$
since the renormalization can be done locally on the target.  Hence it still acts as a differential operator on $F^{\bullet,h}$.  In order to
answer the second point, we recall two key ingredients in the proof of the \v Cech--Dolbeault isomorphism: the $\delb$-Poincar\'e lemma and the
fact that our sheaf is fine.  We now argue that these properties of the sheaf of local operators carry over to the quantum theory.

First of all, we note that the $\delb$-Poincar\'e lemma can be formulated as the vanishing of the higher cohomology groups on a contractible open set $U$, i.e., $H_\delb^q(U,F^{\bullet,h}) = 0$ for~$q > 0$.  We need here the $Q$-Poincar\'e lemma: $H_Q^q(U,F^{\bullet,h}) = 0$ for $q > 0$.  This holds from a general principle which states that perturbative corrections can only destroy, and never create, cohomology classes.%
\footnote{The laplacian $\Delta = Q^*Q + QQ^*$ acts perturbatively as an elliptic operator on a holomorphic vector bundle over a compact
manifold $X$, hence the $Q$-cohomology is isomorphic to $\ker\Delta$ by the Hodge decomposition theorem.  The spectrum of $\Delta$ is discrete
since $X$ is compact, so there is a finite gap between zero and the minimum nonzero eigenvalue.  Although the spectrum is modified by higher
order perturbative corrections, nonzero eigenvalues never go down to zero assuming that the quantum effects are sufficiently small.  On the
other hand, some of the zero eigenvalues may be lifted.  This proves the statement.}
Hence, the cohomology groups which vanish classically therefore continue to vanish in the quantum theory.  This proves the $Q$-Poincar\'e lemma.

The assertion that the sheaf of local operators is fine relies on the existence of a partition of unity.  Suppose we have a ``classical''
partition of unity $\set{\rho_a}$ subordinate to an open cover $\set{U_a}$, so that $\sum_a \rho_a(\phi,\phib) =1$.  Quantum mechanically, the
left-hand side of this equation must be regularized, leading to the renormalized operators $\set{\CNO{\rho_a(\phi,\phib)}}$.  The right-hand
side is of course well defined, thus the renormalization can be done in such a way that $\sum_a \CNO{\rho_a(\phi,\phib)} = 1$ holds.  This gives
a ``quantum'' partition of unity.  Hence, the sheaf of local operators is still fine in the quantum theory.

Given the $Q$-Poincar\'e lemma and that the sheaf of local operators is fine, the isomorphism \eqref{cech-q} follows from the same arguments as
those used to prove the \v Cech--Dolbeault isomorphism.  Therefore, we find that the picture remains the same in quantum perturbation theory:
The cohomology of local operators can be computed from \v Cech cohomology.  The chiral algebra can then be obtained by the nontrivial OPE
between cohomology classes.

Implicit in the argument above is that we have considered a particular local trivialization on $\Sigma$.  Provided that the anomaly cancellation
conditions are satisfied, the correlation functions of the BRST sector are diffeomorphism invariant in $\Sigma$.  Local operators can then be
consistently defined over all of the worldsheet.  This observation allows us to ignore the dependence of the cohomology on $\Sigma$ from our
considerations.

Let us make a final remark: We have actually discussed all the ingredients which are required in the computation of the chiral algebra in
\emph{any} instanton sector $\CM_\beta$.  Nonperturbatively, the sheaf of local operators is defined over $\CM_\beta$.  Since instantons are
nonpropagating, the short distance singularities come only from the fiber direction of the field space (i.e., nonzero modes), and the
renormalization can be done locally on $\CM_\beta$.  Thus, $Q$ induces a well-defined action on the sheaf of local operators.  We can then use
the \v Cech--$Q$ isomorphism to compute the chiral algebra from \v Cech cohomology.

\subsubsection{Gluing Free Theories: Sheaf of Chiral Differential Operators}

Although we have established the \v Cech--$Q$ isomorphism, it is still not quite the end of the story.  To compute the \v Cech cohomology group
$\check H^q(X,\CZ^{0,h})$, we first need to know what the sheaf $\CZ^{0,h}$ represents.  Since $Q$ receives perturbative corrections that depend
on the moduli of the theory such as the target metric and the two-form field, the same is true of the sheaf $\CZ^{0,h}$ of $Q$-closed local
operators.  However, it is in general hard to capture the dependence of $\CZ^{0,h}$ on the moduli.  To avoid this difficulty, we will instead
interpret the moduli as the various ways one can glue fixed \emph{known} spaces patch by patch over the target.  We now make this statement
precise, and explain how one can obtain the $Q$-cohomology using \emph{free theories}.

Let $\set{U_a}$ be a good cover of $X$.  The crucial observation is the following: Since $U_a$ are topologically trivial, one can deform the theory to make it locally free over any given $U_a$.  Specifically, one can deform the metric to be flat and the two-form field to zero locally over $U_a$.%
\footnote{In general, a deformation of the two-form field $T_{ij}$ leads to a different chiral algebra since $T_{ij}$ is not globally defined
over $X$.  However, it is well defined as a local two-form on each $U_a$, which we call $T_a$, if one takes a sufficiently fine open cover.  One
can define a global two-form $\Th$ which coincides with $T_a$ on $U_a$.  One can then subtract $\Th$ from $T_{ij}$ to make the latter zero on
$U_a$.  Since this operation modifies the action by a $Q$-exact term, it does not change the chiral algebra.}
Under each such deformation, $Q$ changes by conjugation as $Q \to \Qt = e^{-\Lambda_a} Q e^{\Lambda_a}$, which leaves the $Q$-cohomology
unchanged.  In terms of the local trivializations, let $\set{f_{ab}}$ be the transition functions of the sheaf $\CF^{q,h}$.  We consider
\begin{equation}
\tilde f_{ab} = \psi_a \circ f_{ab} \circ \psi_b^{-1},
\end{equation}
where $\psi_a$ is given by $\CO \mapsto \widetilde{\CO} = e^{-\Lambda_a} \CO e^{\Lambda_a}$.  The sheaf $\CFt^{q,h}$ with the transition
functions $\set{\tilde f_{ab}}$ is then isomorphic to $\CF^{q,h}$.  Furthermore, let $\Qt_a$ be the restriction of $\Qt$ to $U_a$.  Then, the
collection of operators $\set{\Qt_a}$ glue consistently over $X$ to define a morphism $\Qt\colon \CFt^{q,h} \to \CFt^{q+1,h}$, and from the way
it is defined, the sheaf $\CZt^{q,h}$ of $\Qt$-closed local operators of $\CFt^{q,h}$ is isomorphic to $\CZ^{q,h}$ under the isomorphism
$\CFt^{q,h} \iso \CF^{q,h}$.  In particular, we have the isomorphism of the cohomologies:
\begin{equation}
\check H^q(X,\CZ^{0,h}) \iso \check H^q(X,\CZt^{0,h}).
\end{equation}
Since the action of $\Qt$ is determined locally by a free theory, $\CZt^{q,h}$ consists of $\delb$-closed local sections of $\CFt^{q,h}$.  We
have thus obtained the description of the $Q$-cohomology in terms of free theories.

The question is what transition functions $\set{\tilde f_{ab}}$ one should use.  Of course, the choice of $\set{\tilde f_{ab}}$ is fixed up to
isomorphisms once  the global theory is fixed.  As we mentioned at the beginning of this discussion, the problem is that it is hard to determine
$\CZ^{0,h}$ and to hence obtain the original transition functions $\set{f_{ab}}$ in the first place, since this requires a detailed knowledge of
the perturbative corrections.  Nevertheless, we can start with the local description by known free theories and construct possible global
theories by choosing various transition functions $\set{\tilde f_{ab}}$.

To understand what constraints must be imposed on the choices of $\set{\tilde f_{ab}}$, let us look at how $\CZt^{0,h}$ is described locally.
Sections of $\CZt^{0,h}$ are $\delb$-closed local operators, which do not depend on $\phib$.  In view of the absence of $\phib$, we set $\beta_i
= g_{i\jb}\del_z\phib^\jb$ and $\gamma^i = \phi^i$.  Then, a local operator in $\CZt^{0,h}$ takes a general form:
\begin{equation}
\label{BGO} \CO(\gamma, \del_z\gamma, \del_z^2\gamma, \dotsc; \beta, \del_z\beta, \dotsc).
\end{equation}
Since $\beta$ has dimension $1$, the total number of $\beta$ and $z$-derivatives that appear in \eqref{BGO} is precisely $h$.  The dynamics of
these operators are governed by the free (hence conformally invariant) theory with action
\begin{equation}
S = \frac{1}{2\pi} \int_\Sigma d^2z \beta_i \del_\zb\gamma^i.
\end{equation}
This yields the OPE
\begin{equation}
\beta_i(z) \gamma^j(w) \sim -\frac{\delta_i^j}{z - w},
\end{equation}
with other combinations being regular.  This conformal field theory is called the free $\beta\gamma$ system.  In the mathematical literature,
the sheaf $\CZt^{0,h}$ of local operators of free $\beta\gamma$ systems with this OPE structure is called the \emph{sheaf of chiral differential
operators} (CDOs) \cite{Malikov:1998dw}.

The guiding principle in finding consistent $\set{\tilde f_{ab}}$ is the following.  The path integral measure of the original theory can be
constructed by gluing local measures over $\CM_0 \iso X$ geometrically, such that it preserves the OPEs across patches.  The same must then be
true of the description by locally free theories.  In other words, the transition functions $\set{\tilde f_{ab}}$ must be symmetries of the
conformal field theory.  They are associated with the holomorphic symmetry currents of the free $\beta\gamma$ system.  In dimension $0$, local
sections are holomorphic functions and their OPEs are regular.  Restricted to these operators, $\set{\tilde f_{ab}}$ reduce to the geometrical
transition functions of the underlying manifold~$X$.  In higher dimensions, however, the requirement that the gluing preserves the OPEs leads to
nontrivial transformation laws.

Let us recapitulate what we have discussed so far in this section.  Classically, the chiral algebra is described by Dolbeault cohomology.  One
can exploit the \v Cech--Dolbeault isomorphism to recast the $Q$-cohomology in the language of \v Cech cohomology.  Perturbatively, the action
of $Q$ on local operators receives quantum corrections which can depend intricately on the moduli of the theory.  Even though there is the \v
Cech--$Q$ isomorphism, this itself does not help to make the problem tractable, since perturbative corrections enter the very definitions of the
sheaves that are used in the \v Cech cohomology computations.  To proceed, we note that sheaf theory is local in nature, and locally, the theory
can always be deformed to a free theory.  One can then reconstruct the global theory by ``gluing'' free $\beta\gamma$ systems over the target,
while the moduli is now encoded in the way this gluing is done.

\section{Chiral Algebra with Instantons: $\CP^1$ Model}
\label{CP1}

We saw in the previous section that although there are perturbative quantum corrections to the chiral algebra, the consequences are relatively
minor.  In particular, the grading by charge and scaling dimension of the classical chiral algebra continues to hold in perturbation theory.  As
a result, the perturbative corrections can only destroy some but not all of the classical cohomology classes; for example, the constant operator
$1$ always defines a nontrivial class.  Nonperturbatively, however, we learned in Section~\ref{charge-anomalies} that this preservation of the
grading is no longer true if $c_1(X) \neq 0$.  It is then highly plausible that worldsheet instantons can induce radical modifications to the
chiral algebras of $(0,2)$ models. Our aim in this section is to provide an illuminating example of such a nonperturbative deformation of the
chiral algebra. To this end, we will consider the $(0,2)$ model with target space $X = \CP^1$. We will show that under a nonperturbative
deformation, the structure of the chiral algebra of this simple model will be altered drastically: Worldsheet instantons will destroy $\it{all}$
of the remaining perturbative cohomology classes, thus rendering the chiral algebra trivial.

Recall at this point that quantum corrections can only destroy cohomology classes.  Then, if one wishes to understand the chiral algebra in the
full quantum theory, one must first know the perturbative chiral algebra.  In order to determine the perturbative chiral algebra, one in turn
needs to start with the chiral theory.

The classical chiral algebra for $X = \CP^1$ can be concisely described as follows.  Let $\CO(k)$ be a holomorphic line bundle over $\CP^1$
given by the $k$th power of the hyperplane bundle, so that $TX = \CO(2)$ and $T^*X = \CO(-2)$.  Then, the only nonvanishing Hodge numbers are
$h^0(\CO(k)) = k + 1$ for $k \geq 0$, and $h^1(\CO(k)) = -k - 1$ for $k \leq -2$.  Let us first see how perturbative corrections modify these
classical conditions, using the tools developed in the previous section.

\subsection{The Perturbative Chiral Algebra}

Let $N$ and $S$ be the ``north'' and ``south'' poles of $X = \CP^1$, and let $\set{U_N, U_S}$ be a open cover of $\CP^1$ defined by $U_N = \CP^1
\setminus \set{S}$, \ $U_S = \CP^1 \setminus \set{N}$.  The local holomorphic coordinate on $U_N$ is $\gamma$, and that on $U_S$ is $\gamma' =
1/\gamma$.  On $U_N$, one has the $\beta\gamma$ system described by
\begin{equation}
S = \frac{1}{2\pi} \int_\Sigma d^2z \beta \del_\zb\gamma, \qquad \beta(z) \gamma(w) \sim -\frac{1}{z - w}.
\end{equation}
There is the $\beta'\gamma'$ system on $U_S$ defined similarly.  A nice thing about using free theories is that there is an unambiguous
regularization procedure.  We will renormalize composite operators by the conformal normal ordering, which just throws away the singular part of
the OPE.

As discussed, the transition function of the $\beta \gamma$ system acting on $\gamma$ only, is just the geometrical transition function of
$\CP^1$.  On the other hand, its action on $\beta$ is determined by imposing the condition that the OPEs among the fields $\beta'$ and $\gamma'$
be the same as those among $\beta$ and $\gamma$.  Hence, we have \cite{Malikov:1998dw}
\begin{equation}
\gamma' = 1/\gamma, \qquad \beta'  =  -\gamma^2 \beta + 2\del_z \gamma.
\end{equation}
The term $2\del_z \gamma$ in the expression for $\beta'$ is required by the regularity of the $\beta'\beta'$ OPE.

In the case of $\CP^1$, the chiral algebra splits as  $\CA = \CA^{0,\bullet} \oplus \CA^{1,\bullet}$.  Let us first look at $\CA^{0,\bullet}$.
As vector spaces, $\CA^{0,\bullet}$ is represented by zeroth \v Cech cohomology groups, so the cohomology classes are given by global sections
of the sheaf of CDOs over $\CP^1$.  In dimension $0$, the relevant local operators are holomorphic functions.  Since a holomorphic function on a
compact manifold must be constant, $\CA^{0,0}$ is one-dimensional and generated by the constant operator $1$.

In dimension $1$, we can consider linear combinations of $f(\gamma)\del_z\gamma$ and $f(\gamma)\beta$.  Classically, operators of the former
type are identified as holomorphic one forms in $H^0(\CO(-2))$.  Since this is empty (it is impossible to impose the regularity globally over
$\CP^1$), there is no cohomology class associated.  On the other hand, operators of the latter type are holomorphic vector fields in
$H^0(\CO(2))$.  Classically, we have three such operators; by identifying $\del_\gamma$ with $-\beta$, we can write them as
\begin{equation}
J_- = \beta, \qquad J_3 = -\gamma\beta, \qquad J_+ = -\gamma^2\beta.
\end{equation}
The question is whether they have a quantum extension in the perturbative chiral algebra.  The answer is ``yes.''  Their quantum counterparts
are given by~\cite{Witten:2005px}
\begin{equation}
\begin{split}
J_- &= \beta = -\gamma'^2\beta' + 2\del_z\gamma' , \\
J_3 &= -\gamma\beta = -\gamma'\beta', \\
J_+ &= -\gamma^2\beta + 2\del_z\gamma = \beta'.
\end{split}
\end{equation}
They generate an affine $\mathit{SL}(2)$ algebra at level $-2$:
\begin{equation}
\begin{split}
J_3(z) J_3(w) &\sim -\frac{1}{(z-w)^2}, \\
J_3(z) J_\pm(w) &\sim \pm\frac{J_\pm(w)}{(z-w)^2}, \\
J_+(z) J_-(w) &\sim -\frac{2}{(z-w)^2} + \frac{2J_3}{z-w}.
\end{split}
\end{equation}
The zero mode part generates the $\mathit{SL}(2)$ Lie algebra of the classical theory.  This is a reflection of the underlying geometry of the
target, $\CP^1 \iso SL(2)/B$, where $B$ is the Borel subgroup.

Next, let us consider $\CA^{1,\bullet}$.  In dimension $0$, one already has $h^1(\CO(0)) = 0$ at the classical level.  Thus, $\CA^{1,0}$ is
empty.  In dimension $1$, we can immediately write down one nontrivial cohomology class as follows.  Since $c_1(\CP^1) \neq 0$, the discussion
from Section~\ref{c1} ensures the existence of $\theta$ in $\CA^{1,1}$ in the classical theory.  This is represented by the cocycle
$\set{\theta_{NS}}$, where $2\pi i\theta_{NS} = \del_z\gamma/\gamma$.  Perturbative corrections will not destroy this class.  Since
$h^1(\CO(-2)) = 1$, this is the only class of the form $f(\gamma)\del_z\gamma$.  Operators of the form $f(\gamma)\beta$ do not enter the
cohomology  since $h^1(\CO(2)) = 0$.  Thus, we find that $\CA^{1,1}$ is one-dimensional and generated by $\theta$.

In dimension $2$, the nontrivial cohomology groups have Hodge numbers $h^1(\CO(-2)) = 1$ and $h^1(\CO(-4)) = 3$.  The first case is generated by
the cocycle $\del_z\theta$.  However, it is lifted from the cohomology by perturbative corrections.  To see this, let $T_N = \beta\del_z\gamma$
be the holomorphic stress tensor of the $\beta\gamma$ system, and $T_S = \beta'\del_z\gamma'$ be that of the $\beta'\gamma'$ system.  (We have
omitted the normal-ordering symbol here.) Then, a straightforward computation shows that on $U_N \cap U_S$, one has
\begin{equation}
\label{TNTS} T_N - T_S = 2\pi i\del_z\theta_{NS}.
\end{equation}
That is, the cocycle $\del_z\theta$ vanishes in the \v Cech cohomology.

In order to find the sigma model counterpart of \eqref{TNTS}, we proceed in a similar manner as we did to find the representation $\theta$.  In
the case at hand, $T_N$ and $T_S$ must correspond, under the isomorphism from the global theory to the local $\beta\gamma$ systems, to the local
descriptions of $T_{zz}$ in $U_N$ and $U_S$.  Then, via the \v Cech--$Q$ isomorphism, the counterpart of \eqref{TNTS} in the global description
will be $[Q,T_{zz}] = 2\pi i\del_z\theta$.  Thus, $\del_z \theta$ is $Q$-exact.  Plugging in the classical expression of $\theta$ derived in
Section~3.1 into this equation, we obtain (ignoring higher order corrections)
\begin{equation}
[Q,T_{zz}] = \del_z(R_{i\jb} \del_z\phi^i \alpha^\jb).
\end{equation}
In fact, this relation between $T_{zz}$ and $\del_z\theta$ holds for any target space with $c_1(X) \neq 0$ (see \cite{Nekrasov:2005wg},
Section~2.4.5).  In short, the chiral algebra lacks the holomorphic stress tensor whenever $c_1(X) \neq 0$.  This is consistent with the
discussion in Section~3.2.

The other cohomology classes in $\CA^{1,2}$ are easy to find.  Here, the Serre duality $H^1(X,\CO(-4)) = H^0(X,\CO(2))^*$ is rather indicative.
Using the perturbative cohomology classes we have found already, we can construct new ones: $J_\pm\theta$, $J_3\theta$.  Thus, we find the
isomorphisms $\CA^{0,0} \iso \CA^{1,1}$ and $\CA^{0,1} \iso \CA^{1,2}$.  In fact, it can be shown \cite{Malikov:1998dw} that the isomorphism
$\CA^{0,h} \iso \CA^{1,h+1}$ persists in higher dimensions $h \geq 2$ as well.

\subsection{The Nonperturbative Chiral Algebra}

We have seen above that in the lowest dimension of the perturbative chiral algebra of the $\CP^1$ model, the bosonic (charge $0$) and fermionic
(charge $1$) subspaces are generated by the cohomology classes $1$ and $\theta$, respectively.  Furthermore, cohomology classes of higher
dimensions are constructed by acting ``creation'' cohomology classes---which are classes in $\CA^{0,\bullet}$ that are hence bosonic---on the
``ground'' cohomology classes $1$ and $\theta$.  As we will elaborate in Section~5, there is a striking resemblance between the way the
perturbative chiral algebra of the $\CP^1$ model is built up, and the way the Fock space of states in the associated closed string on $X =
\CP^1$ is generated.

Given the above analogy between physical states and perturbative cohomology classes, one suspects that instantons may connect these cohomology
classes in a nonperturbative relation such as $\{Q, \theta\} \sim 1$.  Even though $1$ and $\theta$ differ by one in dimension, this is possible
since instantons violate the scaling dimensions of operators as we explained in Section~\ref{anomalies}.  If such a relation exists, it will
imply that the chiral algebra of this model would be completely trivialized---the identity operator $1$ will be $Q$-exact, and it will act on
any operator to make it $Q$-exact as well.  Let us now ascertain if the presence of worldsheet instantons leads to such a dramatic,
nonperturbative phenomenon.

\subsubsection{The One-Instanton Computation}

The supercharge is given by $Q = \oint J d\zb$, where $J = g_{i\jb} \del_\zb\phi^i \alpha^\jb$.  Then, the commutator $\{Q, \theta\}$ will be
nonvanishing only if the OPE $J(\zb) \cdot \theta(w,\wb)$ in an instanton background contains an \emph{antiholomorphic single pole} as $\zb \to
\wb$.  We need to specify the geometry of the worldsheet in order to compute this OPE.  Since we wish to study the instanton effects on a
nonanomalous $\CP^1$ model, we will consider the theory defined on a worldsheet with trivial canonical line bundle.  The physical and twisted
models are then the same if we choose a trivial spin structure, since the twisting does nothing in this case.  Hence, let us just focus on the
physical model.

Interesting examples are $\Sigma = \C$ or $S^1 \times \R$.  We will restrict ourselves for the moment to instantons of degree one, and consider
the lowest order in perturbation theory around them.  The general case will be treated in the next subsection.  To this order, the relevant OPEs
are computed with respect to a free conformal field theory.  Thus, one is free to conformally map the infinitely long cylinder and the complex
plane worldsheets to the sphere by adding points at infinity and inserting vertex operators.  Since we are considering the physical model, there
is no anomaly on the sphere.  Instantons are then one-to-one holomorphic maps from $\Sigma = \CP^1$ to $X = \CP^1$, i.e., automorphisms
of~$\CP^1$.

With such an interpretation of instantons at hand, we can now proceed to determine the $R$ charge anomaly, which gives us the required condition for the nonvanishing of the correlation function.  As discussed in Section \ref{charge-anomalies}, the charge anomaly for physical models is given by $\int_{\Sigma} \phi^*c_1(X)$.  For $X = \CP^1$ and degree one instantons, this is~$2$.  Hence, for the correlation function to be \textit{a priori} nonvanishing, it must contain exactly two $\alpha$~zero modes so that they can soak up the zero modes present in the path-integral measure.%
\footnote{More precisely, the index which quantifies the charge anomaly is given by the difference in the number of $\alpha$ and $\rho$ zero
modes.  However, there are no $\rho$ zero modes in the present case.}
We wish to compute the singular part of the correlation function
\begin{equation}
\label{I}
I = \bigvev{J(\zb) \theta(w, \wb)}
\end{equation}
in the one-instanton background.  Since $J$ and $\theta$ each has one $\alpha$ field, this in general does not vanish.  Note that the integration over the fermionic nonzero modes yields the fermionic determinant $\Det'\Delta_F$, which is just a constant after an appropriate regularization.  Similarly, the integration over the instanton moduli space does not produce short distance singularities either, since instantons are non-propagating.  Hence, the relevant term in $I$ comes from the integration over the bosonic nonzero modes.

For an $\CN = (0,2)$ supersymmetric sigma model on a K\"ahler manifold, the two-form field $T_{ij} = 0$.  The action is then given by
\begin{equation}
S= \frac{1}{2\pi} \int_{\Sigma} d^2z \, g_{i\jb}(\phi,\phib)
   \bigl(\del_\zb\phi^i \del_z\phi^\jb + \rho^i D_z\alpha^\jb\bigr) + S_B.
\end{equation}
Expanding the bosonic action $\phi = \phi_0 + \varphi$ around an instanton $\phi_0$ and keeping only the quadratic terms, one obtains
\begin{equation}
S_0 = \int_{\Sigma} d^2z \, g_{i\jb}(\phi_0,\phib_0)
      \del_\zb\vphi^i \del_z\vphi^\jb.
\end{equation}
Although instantons are global on the target, the fluctuations $\varphi$ are local in nature.  Hence, the correlation function $I$ can be computed using the following OPE derived from the above free field action $S_0$:
\begin{equation}
\label{free OPE}
\vphi^i(z,\zb) \vphi^\jb(w,\wb)
\sim -g^{i\jb}\bigl(\phi_0(w),\phib_0(\wb)\bigr) \ln|z-w|^2.
\end{equation}
Recall at this point that we are looking specifically for antiholomorphic single poles in~$I$.  From the explicit forms of $J$ and $\theta$, we have
\begin{equation}
\label{I}
I = \bigvev{\bigl(g_{i\jb} \del_\zb\phi^i \alpha^\jb\bigr)(\zb)
            \bigl(g_{k\lb} \del_z\phi^k \alpha^\lb\bigr)(w,\wb)}
\end{equation}
Since the zeroth order of the expansion of $\alpha(\zb) \alpha(\wb)$ around $\wb$ vanishes by Fermi statistics, we must find contractions in~$I$ that produce antiholomorphic \emph{double} poles.  Seemingly, there are none.

However, there is a fine subtlety here.  We saw in Section~\ref{localization} that $\alpha$ can be expanded in eigenmodes of the laplacian $\Delta_F$ as
\begin{equation}
\alpha(z,\zb; \phi) = \sum_r c_0^r \ub_{0,r}(z,\zb; \phi) + \sum_n c^n \ub_n(z,\zb; \phi).
\end{equation}
As noted there, these eigenmodes contain fluctuating bosonic modes, and therefore, they can produce short distance singularities upon interaction
with other fluctuating fields.

For this reason, we must examine more carefully the dependence of the fermionic zero modes on the bosonic field.  Let us recall how we could parametrize the field space near instantons.  A point $\phi$ in the neighborhood $\CM'$ of the instanton moduli space $\CM \subset \Map(\Sigma, X)$ can be identified with a vector at a nearest point $\phi_0 \in \CM$.  This vector can be expanded in the eigenmodes as
\begin{equation}
\vphi(z,\zb; \phi_0) = \sum_n a^n u_n(z,\zb;\phi_0), \qquad \vphib(z,\zb; \phi_0) = \sum_n \ab^n \ub_n(z,\zb;\phi_0).
\end{equation}
Together with the complex parameters $\{\zeta_0^r\}$ of instantons, the coefficients $\{a^n\}$ define holomorphic coordinates on $\CM'$.  We
will write the full dependence of the bosonic field as $\phi^i(z,\zb; \zeta, a)$.  By construction, the dependence of an instanton is
$\phi_0^i(z;\zeta)$.  Via the identification $\Gamma(\phi^*\TXb) \iso \overline{T_\phi\CM'}$, we may view $\alpha(z,\zb; \phi)$ as an odd vector
in $\overline{T_\phi\CM'}$.  Thus, it can be expanded as
\begin{equation}
\alpha(z,\zb; \phi) = \sum_r \tilde c_0^r \del/\del\zetab_0^r + \sum_n \tilde c^n \del/\del \ab^n.
\end{equation}
As vector fields, $\del/\del\zetab_0^r$, $\del/\del\ab^n \in \overline{T_\phi\CM'}$ have components $\del\phi^\ib/\del\zetab_0^r$ and
$\del\phi^\ib/\del\ab^n$, respectively.  On $\CM$ they reduce to $\ub_{0,r}$ and $\ub_n$, and also, the coefficients $\tilde c_{0,r}$, $\tilde
c_n$ reduce to $c_{0,r}$ and $c_n$.

A one-instanton maps the worldsheet to a two-cycle in the target in a one-to-one manner.  Hence, we may also specify the bosonic field as
$\phih^i(\phi_0,\phib_0; \zeta, a)$, which is related to the previous presentation by
\begin{equation}
\phi^i(z,\zb; \zeta, a) = \phih^i\bigl(\phi_0(z;\zeta),\phib_0(\zb;\zetab); \zeta, a\bigr).
\end{equation}
Then, the expansion of $\alpha$ becomes
\begin{equation}
\label{EXPALPHA} \alpha^\ib(z,\zb; \phi) = \sum_r \tilde c_0^r
  \Bigl(\frac{\del\phih^\ib}{\del\phi_0^\jb}
        \frac{\del\phi_0^\jb}{\del\zetab_0^r}
        + \frac{\del\phih^\ib}{\del\zetab_0^r}\Bigr)
  + \sum_n \tilde c^n \frac{\del\phih^\ib}{\del\ab^n}.
\end{equation}
But since $\del_\zb\phi^\ib = \del_\zb\phi_0^\jb \del\phih^\ib/\del\phi_0^\jb$, we see that the zero mode part of $\alpha$ contains%
\footnote{The other terms in the expansion \eqref{EXPALPHA} do not contribute at the end of the computation.  As such, we will omit their
discussion.}
\begin{equation}
\sum_r c_0^r \frac{\del_\zb\phi^\ib}{\del_\zb\phi_0^\jb} \ub_r^\jb(\zb; \phi_0).
\end{equation}
Clearly, $\del_\zb\phi^\ib$ in the $\alpha$ zero mode in $\theta$ can now contract with $\del_\zb\phi^i$ in $J$, and produce an antiholomorphic
double pole; indeed, we find that the integration over the bosonic nonzero modes in $I$ gives the term
\begin{equation}
\label{JTHETA2}
\frac{1}{(\zb - \wb)^2}
\Bigl(g_{i\jb} \del_w\phi_0^i
      \frac{1}{\del_\wb\phi_0^\kb}
\Bigr)(w,\wb)
\sum_{r \neq s} c_0^r c_0^s
\ub_{0,r}^\jb(\zb; \phi_0)
\ub_{0,s}^\kb(\wb; \phi_0).
\end{equation}
Above, we have omitted for clarity of discussion, higher order terms that are negligible in the large volume limit, and those terms that give rise to $Q$-exact quantities in the final computation.

The rest of the computation goes as in \cite{Dine:1987bq, Distler:1987ee, Berglund:1995yu}.  An instanton of degree one, which is an
automorphism of $\CP^1$, is given by
\begin{equation}
\label{mobius} \phi_0(z) = \frac{az + b}{cz +d};
\end{equation}
where $a$, $b$, $c$, $d$ are complex numbers with $ad - bc =1$.  One-instantons thus form a three-dimensional moduli space $\CM_1$.  The $\alpha$ zero modes can be obtained from the superconformal variations of ${\bar \phi}_0$:
\begin{equation}
\label{zero-mode part} \ub_{0,1}(\zb) = \frac{1}{\bar c\zb + \bar d}, \qquad \ub_{0,2}(\zb) = \frac{1}{\bar c(\bar c\zb + \bar d)^2}.
\end{equation}
Now, note that the correlation function~$I$ is computed as
\begin{equation}
I = \int d\CM_1 dc_0^1 dc_0^2 \, \CD\mu' e^{-S} J(\zb) \theta(w,\wb),
\end{equation}
where $d\CM_1$ is the volume form of $\CM_1$, and $\CD\mu'$ is the measure for nonzero modes.  Plugging \eqref{JTHETA2} in for the bosonic nonzero mode integration, and performing the integration over the $\alpha$ zero modes and fermionic nonzero modes, we obtain
\begin{equation}
\begin{split}
\bigvev{J(\zb) \theta(w, \wb)}
&= e^{-t} \Det'\Delta_F
      \int_{\CM_1} d\CM_1 \, \frac{1}{\zb - \wb}
      \bigl(g_{i\jb} \del_w\phi_0^i\bigr)(w,\wb)
      \frac{1}{(\cb\zb + \db)^2} \\
&= e^{-t} \Det'\Delta_F
   \int_{\CM_1} d\CM_1 \, \frac{1}{\zb - \wb}
   \bigl(g_{i\jb} \del_w\phi_0^i\bigr)(w,\wb) \del_\zb\phib_0^\jb(\zb),
\end{split}
\end{equation}
where $t = S_B$ for one-instantons.  Performing the contour integral over $\zb$ around $\wb$, this becomes
\begin{equation}
\bigvev{\{Q,\theta\}(w, \wb)}
= e^{-t} \Det'\Delta_F
  \int_{\CM_1} d\CM_1
  \bigl(g_{i\jb} \del_w\phi_0^i \del_\wb\phib_0^\jb\bigr)(w,\wb).
\end{equation}
Now, we can trade the integration above over one of the parameters of the instantons with the integration over the location $(w,\wb)$ of $\theta$.  This yields
\begin{equation}
\bigvev{\{Q,\theta\}(w, \wb)}
\propto e^{-t} \Det'\Delta_F \int_\Sigma \phi_0^* \, \omega,
\end{equation}
where $\omega$ is the K\"ahler form.  The right-hand side is just the partition function in the zero-instanton sector $\bigvev{1}_0$, multiplied by the instanton factor $e^{-t}$.  Therefore, we indeed have the relation $\{Q, \theta\} \sim 1$ in the one-instanton background.

\subsubsection{Trivialization of the Chiral Algebra}

We have shown above that there is a relation $\{Q, \theta\} \sim 1$ induced by instantons of degree one, which implies that $\CO \sim
\{Q,\CO\theta\}$ for any $Q$-closed operators $\CO$.  Hence, the chiral algebra is manifestly trivialized in the presence of one-instantons.
Since in general quantum corrections cannot create cohomology classes, the chiral algebra should remain trivial in the full quantum
theory---where one considers instantons of all degrees,  not just those of degree one---if it is already trivialized by one-instantons.  We now
argue that this is indeed the case.

The idea is to construct corrections to a $Q$-exact operator $\CO$ order by order in accordance with the degree of instantons, and show that the
corrected operator, although $Q$-closed under the instanton-corrected action of $Q$, is again $Q$-exact at each order.  Let us begin by writing
the supercharge as $Q = Q_0 + e^{-t} Q_1 + e^{-2t} Q_2 + \dotsb$.  We write $Q^{(k)} = \sum_{n=0}^k e^{-nt} Q_i$.  What we want to show is that
if the $Q^{(k)}$-cohomology is trivial, then the $Q^{(k+1)}$-cohomology is also trivial.  Below we illustrate how this works in the case of $k =
0$.

Let $\CO_0$ be a $Q_0$-closed operator.  Since the $Q^{(0)}$-cohomology is trivial by assumption, $\CO_0 = \{Q_0,V_0\}$ for some~$V_0$.  After
the correction $Q_1$ is included, $\CO_0$ will in general not be $Q$-closed any longer.  Suppose we can find a correction $\CO_1$ so that $\{Q_0
+ e^{-t} Q_1, \CO_0 + e^{-t} \CO_1\} = 0$.  The correction~$\CO_1$ must satisfy the equation
\begin{equation}
\{Q_0,\CO_1\} + \{Q_1,\CO_0\} = \{Q_0,\CO_1\} + \{Q_1,\{Q_0,V_0\}\} = 0.
\end{equation}
On the other hand, from the required nilpotency of $Q^{(1)} = Q_0 + e^{-t} Q_1$, we obtain $\{Q_0,Q_1\} = 0$.  Using this, the above equation
becomes $\{Q_0, \CO_1 - \{Q_1,V_0\}\} = 0$.  Thus, $\CO_1 - \{Q_1,V_0\}$ defines a $Q^{(0)}$-cohomology class, and again by assumption, there
exists $V_1$ such that $\CO_1 - \{Q_1,V_0\} = \{Q_0,V_1\}$.  It then follows that
\begin{equation}
\CO_0 + e^{-t}\CO_1 = \{Q_0 + e^{-t} Q_1, V_0 + e^{-t} V_1\}.
\end{equation}
We thus conclude that any $Q^{(1)}$-closed operator $\CO_0 + e^{-t}\CO_1$ must actually be $Q^{(1)}$-exact, i.e., the $Q^{(1)}$-cohomology is
trivial.

In our case, the $Q$-cohomology becomes empty after the one-instanton corrections are taken into account.  The one-instanton computation
captures $Q^{(1)}$.  Starting from this trivial $Q^{(1)}$-cohomology, one can easily show that the entire chiral algebra is trivial in the full
quantum theory by repeatedly applying the same argument as above.

A number of interesting observations follow from this trivialization of the chiral algebra.  One of them is that the generating series of the
chiral algebra
\begin{equation}
\CA(q)  = q^{-d/12} \sum_{h = 0}^\infty
           \bigl(q^h \Tr_{\CA^{\bullet,h}} (-1)^F\bigr)
\end{equation}
vanishes nonperturbatively, even though it is nonzero in perturbation theory due to a shift in dimension between $\CA^{0,\bullet}$ and
$\CA^{1,\bullet}$ \cite{Malikov:1998dw}.  This should be contrasted to the elliptic genus, which is defined with respect the cohomology of
states and receives no quantum corrections either perturbatively or nonperturbatively.

Another point to note is that the local operators $J_\pm$, $J_3$ are automatically lifted out of the supersymmetric spectrum.  As discussed
earlier in the previous section, these operators are intimately related to the geometry of $X = \CP^1$.  Therefore, one can in some sense
interpret this as a ``quantum'' deformation of the geometry of the target space by worldsheet instantons.

\subsubsection{Supersymmetry Breaking}

The trivialization of the chiral algebra also results in nontrivial consequences for the supersymmetric spectrum.  Let $\ket{\Psi}$ be a
$Q$-closed state, i.e., $Q\ket{\Psi} = 0$.  In the presence of worldsheet instantons, we have
\begin{equation}
1 \cdot \ket{\Psi} \sim \{Q,\theta\}\ket{\Psi}
   = Q\bigl(\theta\ket{\Psi'}\bigr).
\end{equation}
In other words, all $Q$-closed states are actually $Q$-exact.  This means that there are no supersymmetric states in the full quantum theory.
Perturbative supersymmetric states, which are characterized by the condition $\Lb_0 = 0$, are lifted by instanton effects, leading to a
spontaneous breaking of supersymmetry.  Let us now turn to a closer look at this phenomenon from the canonical quantization viewpoint.

\section{Holomorphic Morse Theory on Loop Space}

In the context of supersymmetric quantum mechanics, Morse theory provides a powerful tool to analyze the nonperturbative supersymmetric spectrum
\cite{Witten:1982im}.  To adapt this approach to our case, we must make two generalizations.  First, our models are quantum field theories in
two dimensions, so we must consider Morse theory on loop space along the lines of Floer \cite{F}.  Second, since we have supercharges only in
the right-moving sector, we must consider holomorphic Morse theory \cite{W2}.  As a result, the supersymmetric spectrum is less constrained (in
particular, it is infinite-dimensional), and the applicability of Morse theory is severely restricted.  Nevertheless, as we will demonstrate,
holomorphic Morse theory on loop space will prove to be fruitful, as it will provide us with further insights into the physics that underlie the
spontaneous breaking of supersymmetry in the $\CP^1$ model.  Moreover, we will find that under a ``quantum'' deformation of the target geometry,
the kernels of certain twisted Dirac operators on $\CP^1$ will become trivial.  An interesting attempt to study quantum mechanical models from
this perspective has also been undertaken by Frenkel et al.~\cite{Frenkel:2006fy}.

\subsection{Canonical Quantization of $(0,2)$ Models}

Since our discussion will revolve around physical states, it will be most illuminating to proceed with the canonical quantization viewpoint.
Doing so will lead us naturally to supersymmetric quantum mechanics on loop space.  But first, let us review how states emerge in the context of
path integrals.

Suppose that one punctures a small hole $D^2$ in the worldsheet $\Sigma$.  To compute the path integral with this worldsheet, one has to specify
the boundary condition $\Phi|_{S^1}$ on the boundary circle $S^1 = \del D^2$, where we have written all the local data (i.e., fields and their
derivatives) collectively as $\Phi$.  More generally, one may assign a weight $\Psi(\Phi|_{S^1})$ to each boundary condition $\Phi|_{S^1}$, so
that the correlation function is given by
\begin{equation}
\label{CFPSI} \bigvev{\prod_a \CO_a}_\Psi = \int \CD\Phi \, e^{-S(\Phi)} \Psi(\Phi|_{S^1}) \prod_a \CO_a(\Phi).
\end{equation}
In this expression, we have explicitly indicated the boundary condition used in the path integral.  The wavefunction $\Psi$ defines a state
$\ket{\Psi}$.

Symmetries of the theory induce transformations of states as they change the boundary conditions.  Suppose a symmetry transformation $U$ acts on
fields by $\Phi \to \Phi^U$, under which the action and the path integral measure are left invariant, i.e., $S(\Phi) = S(\Phi^U)$ and $\CD\Phi =
\CD\Phi^U$.  By performing a change of integration variables, we see that the correlation function \eqref{CFPSI} satisfies
\begin{equation}
\int \CD\Phi \, e^{-S(\Phi)} \Psi(\Phi|_{S^1}) \prod_a \CO_a(\Phi) = \int \CD\Phi \, e^{-S(\Phi)} \Psi(\Phi^U|_{S^1}) \prod_a \CO_a(\Phi^U).
\end{equation}
If we now define a new state $\Psi^U$ and operators $\CO_a^U$ by $\Psi^U(\Phi|_{S^1}) = \Psi(\Phi^U|_{S^1})$ and $\CO_a^U(\Phi) =
\CO_a(\Phi^U)$, we obtain
\begin{equation}
\bigvev{\prod_a \CO_a}_\Psi = \bigvev{\prod_a {\CO_a^U}}_{\Psi^U}.
\end{equation}
Hence the transformation $\Psi \to \Psi^U$ and $\CO_a \to \CO_a^U$ leaves the correlation function invariant.  The former thus gives the action
of $U$ on states, while the latter is the transformation of operators which appears also for closed worldsheets.  If $U$ is a continuous
symmetry, such is the case for supersymmetry, its generators act as differential operators on the space of states.  We will see this explicitly
in a moment.

To make a transition to supersymmetric quantum mechanics, we need to introduce a time direction.  Thus, let us take the worldsheet to be an
infinitely long cylinder $\Sigma = S^1 \times \R$ with a flat metric $ds^2 = d\sigma^2 + d\tau^2$, where $\sigma = \sigma + 2\pi$.  This is
topologically equivalent to a $\CP^1$ with two points removed.  In this case, we have two boundaries on~$\Sigma$, one for each end of the
cylinder at infinity.  The action diverges and hence amplitudes vanish unless $\|\del_\tau\phi\| \to 0$ as $\tau \to \pm\infty$.  The boundary
conditions for $\phi(\sigma,\tau)$ are then specified by its asymptotic behavior, i.e., the loops $\phi(\sigma,\pm\infty)$.  In the purely
bosonic case, a state~$\ket{\Psi}$ is thus a function on the loop space $LX = \Map(S^1,X)$.

In our supersymmetric case, we must also consider the fermions.  For the boundary states to be invariant under supersymmetry, the fermions, like
the bosons, must depend only on $\sigma$ as $\tau \to \pm\infty$.  Thus, their boundary conditions are specified by $\alpha(\sigma,\pm\infty)$
and~$\rho(\sigma,\pm\infty)$.  The fields at infinity define an infinite-dimensional supermanifold, which has $LX$ as its underlying manifold.
States are functions on this supermanifold; as we will see, we may interpret them as spinors on~$LX$.  In this sense, our theory is
supersymmetric quantum mechanics on $LX$.  In the Schr\"odinger picture, the fields at time~$\tau$ define a state which is a spinor at
$\phi(\blank,\tau) \in LX$ given by $\alpha(\blank,\tau)$ and $\rho(\blank,\tau)$.  Once a field configuration is specified over the cylinder,
it describes a complete evolution of the initial state at $\tau = -\infty$ that ends up with the final state at $\tau = +\infty$, with the
amplitude weighted by $e^{-S}$.

Let us now canonically quantize the theory.  We will assume that the target space $X$ is K\"ahler, and set the $B$-field equal to the K\"ahler form $\omega = (i/2) g_{i\jb} d\phi^i \wedge d\phi^\jb$.  Then, our action will be given by%
\footnote{We make the replacement $\rho \to -i\rho$ to simplify formulas.  Then, $\{Q,\rho^i\} = -i\del_\zb\phi^i$.}
\begin{equation}
\begin{split}
S &= \frac{1}{2\pi} \int_\Sigma d^2z
     \bigl\{Q, ig_{i\jb} \rho^i \del_z\phi^\jb\bigr\}
     + \frac{1}{2\pi} \int_\Sigma \phi^*\omega \\
  &= \frac{1}{2\pi} \int_\Sigma d\sigma d\tau
     \Bigl(\frac{1}{2} g_{i\jb}
     (\del_\tau\phi^i \del_\tau\phi^\jb + \del_\sigma\phi^i \del_\sigma\phi^\jb)
     - \rho_\ab (D_\tau + iD_\sigma) \alpha^\ab\Bigr).
\end{split}
\end{equation}
where $z = \sigma + i\tau$, \ $d^2z = idz\wedge d\zb$, and the covariant derivatives are taken with respect to the spin connection.  Upon
quantization,  the fermions obey the anticommutation relation
\begin{equation}
\label{CAR} \{\rho_\ab(\sigma,\tau), \alpha^\bb(\sigma',\tau)\} = 2\pi \delta_\ab^\bb \delta(\sigma - \sigma').
\end{equation}
This is a loop-space analog of the Clifford algebra.  On the other hand, the quantization of the bosonic field identifies the functional
derivative on $LX$ as
\begin{equation}
\label{CCR} \frac{\delta}{\delta\phi^i}
   = \frac{1}{4\pi} g_{i\jb} \del_\tau\phi^\jb, \qquad
\frac{\delta}{\delta\phi^\ib}
   = \frac{1}{4\pi} g_{\ib j} \del_\tau\phi^j
     + \frac{1}{2\pi} \omega_\ib{}^\ab{}_\bb \rho_\ab \alpha^\bb.
\end{equation}
Note that the second equation is free of the ambiguity associated with the operator ordering, thanks to the antisymmetry of the spin connection.

\subsection{Perturbative Supersymmetric Spectrum}
\label{PSS}

We now turn to the analysis of the supersymmetric spectrum.  In the $\CN = (0,2)$ case, supersymmetric states must be annihilated by $Q$ and
$Q^\dagger$.  Combined with unitarity, it follows that the supersymmetric spectrum consists of states which satisfy $\{Q,Q^\dagger\} = (H - P)/2
= \Lb_0 = 0$.  As $P$ is unbounded, the space of supersymmetric states is in general infinite-dimensional.

Since the supersymmetric spectrum is the harmonic space of $Q$, it is also isomorphic to the $Q$-cohomology of states, where each cohomology
class has a unique representative given by a supersymmetric state.  To determine this spectrum of states, it will be useful to introduce
operators
\begin{equation}
\label{Q_0}
\begin{split}
Q_0 &= \frac{1}{4\pi} \int_{S^1}d\sigma g_{\ib j}\alpha^\ib \del_\tau\phi^j
     = \int_{S^1}d\sigma \alpha^\ib \frac{D}{D\phi^\ib}, \\
Q_0^\dagger
    &= -\frac{1}{4\pi} \int_{S^1}d\sigma g_{i\jb}\rho^i \del_\tau\phi^\jb
     = -\int_{S^1}d\sigma \rho^i \frac{\delta}{\delta\phi^i}.
\end{split}
\end{equation}
Here the covariant functional derivative is given by
\begin{equation}
\frac{D}{D\phi^\ib} = \frac{\delta}{\delta\phi^\ib}
  + \frac{1}{2\pi} \omega_\ib{}^\ab{}_\bb \alpha^\bb \rho_\ab.
\end{equation}
In view of the anticommutation relation \eqref{CAR}, this is a loop-space generalization of the covariant derivative.  One may consider $Q_0$ as
``half'' the Dirac operator on $LX$, while $Q_0^\dagger$ being the other half.

The point of introducing such operators is that they can be deformed to the actual supercharges without altering the cohomology.  Let $h$ be a
function on $LX$.  We consider the conjugation
\begin{equation}
\label{Qs}
\begin{split}
Q_s &= e^{sh/2\pi} Q_0 e^{-sh/2\pi}
     = Q_0 - \frac{s}{2\pi} \int_{S^1} d\sigma
             \alpha^\ib \frac{\delta h}{\delta\phi^\ib}, \\
Q_s^\dagger
    &= e^{-sh/2\pi} Q_0^\dagger e^{sh/2\pi}
     = Q_0^\dagger - \frac{s}{2\pi} \int_{S^1} d\sigma
                     \rho^i \frac{\delta h}{\delta\phi^i}.
\end{split}
\end{equation}
These are supercharges of the theory defined by the action
\begin{equation}
\label{S_h}
\begin{split}
S_h
  &= \frac{1}{2\pi} \int_\Sigma d^2z
     \Bigl\{Q_s, \rho^i \Bigl(\frac{1}{2} g_{i\jb} \del_\tau\phi^\jb
                              - s\frac{\delta h}{\delta\phi^i}\Bigr)\Bigr\}
     + \frac{s}{2\pi} \int_{-\infty}^\infty d\tau \del_\tau h \\
  &= \frac{1}{2\pi} \int_\Sigma d\sigma d\tau
     \Bigl(\frac{1}{2} g_{i\jb} \del_\tau\phi^i \del_\tau\phi^\jb
           + 2s^2 g^{i\jb}
           \frac{\delta h}{\delta\phi^i} \frac{\delta h}{\delta\phi^\jb}
           + \text{fermions}\Bigr).
\end{split}
\end{equation}
The function $h$ in our present context can be described as follows \cite{Hori:2003ic}.  Given a loop $\gamma \in LX$, let $\gamma_u\colon [0,1]
\to LX$ be a homotopy connecting $\gamma_0$ and $\gamma_1 = \gamma$, where $\gamma_0$ is a reference loop in the component of $LX$ in which
$\gamma$ lies.  Such a homotopy defines a map $\gammah$ from the annulus $A = [0,1]\times S^1$ to $X$.  Then, $h(\gamma)$ is the pullback of the
K\"ahler form by $\gammah$:
\begin{equation}
\label{MLX} h(\gamma) = \int_A \gammah^*\omega.
\end{equation}
Under a variation $\delta\gamma$ of the loop, $h$ changes by
\begin{equation}
\label{deltah} \delta h = \frac{i}{2} \int_{S^1} d\sigma
  \bigl(-g_{i\jb}\delta\gamma^i\del_\sigma\gamma^\jb
        + g_{i\jb}\del_\sigma\gamma^i\delta\gamma^\jb\bigr).
\end{equation}
Plugging this in \eqref{Qs}, we obtain
\begin{equation}
\label{QsEXPLICIT}
\begin{split}
Q_s &= \frac{1}{4\pi} \int_{S^1}d\sigma
       g_{\ib j}\alpha^\ib (\del_\tau - is\del_\sigma)\phi^j, \\
Q_s^\dagger
    &= -\frac{1}{4\pi} \int_{S^1}d\sigma
        g_{i\jb}\rho^i (\del_\tau - is\del_\sigma)\phi^\jb.
\end{split}
\end{equation}
From this expression, we see that the deformed supercharges reduce to the supercharges of the original theory when $s = 1$.   Since  the
cohomology remains unchanged under conjugation, we find that the $Q$-cohomology is isomorphic to the $Q_0$-cohomology, which is a spinor
cohomology on~$LX$.

The structure of the harmonic space simplifies considerably if we work with the deformed supercharges.  Note that $\Delta_s =
\{Q_s,Q_s^\dagger\}$ contains a ``potential'' proportional to~$s^2$:
\begin{equation}
\frac{s^2}{2\pi} \int_{S^1} d\sigma \Bigl\|\frac{\delta h}{\delta\phi}\Bigr\|^2.
\end{equation}
Hence in the limit $s \to \infty$, supersymmetric states must localize around the critical points of $h$ in $LX$.  In the present case, they are
solutions of the equation $\del_\sigma\phi = 0$, i.e.,~constant loops.  Intuitively, this can be understood as follows.  From
\eqref{QsEXPLICIT}, we see that $Q_s$ is the supercharge of the theory defined on the worldsheet with coordinates $(\sigma_s, \tau)$ and metric
$d\sigma_s^2 + d\tau^2$, where $\sigma_s = \sigma/s$.  The circumference of the cylinder is $2\pi/s$.  As $s$ gets larger and larger, the
cylinder becomes narrower and narrower.  It then takes an increasing amount of energy to ``stretch'' the loop to make a large circle in the
target space.

Since tiny loops cannot ``feel'' the curvature of the target space, the harmonic space may be approximated by the Fock space of free closed
strings.  One can then systematically construct the harmonic space order by order in $1/s$.  Looking back at the expressions~\eqref{QsEXPLICIT}
of the deformed supercharges, we notice that the parameter~$s$ appears only in the combination $sg_{i\jb}$.  Thus, as far as the supersymmetric
spectrum is concerned, the expansion in $1/s$ is equivalent to sigma model perturbation theory.  The widths of localized supersymmetric states
are controlled by $1/s$, and by increasing~$s$ we are effectively inflating the target space.

To construct the Fock space, we first expand the fields in the eigenmodes of $iD_\sigma$.  At each $\tau$, let $\phi_0(\tau)$ be the center of
the loop $\phi(\blank,\tau) \in LX$ defined by the zero mode of the Fourier expansion in a given local trivialization.  Using geodesic
coordinates, a fluctuation of the loop around the constant loop $\phi_0(\tau) \in LX$ can be identified with a vector field
$\varphi(\sigma,\tau)$ over $S^1 \times \{\tau\}$ valued in $T_{\phi_0(\tau)}X$.  The fermionic fields can also be considered as vector fields
at the center.  Pulled back to $S^1 \times \{\tau\}$ by the constant loop $\phi_0(\tau)$, the covariant derivative $iD_\sigma$ reduces to
$i\del_\sigma$.  Thus, the mode expansions are given by
\begin{equation}
\begin{alignedat}{3}
\varphi^i(\sigma,\tau)
  &=  s^{-1/2} \sum_{n \neq 0} \phi^i_n(\tau) e^{in\sigma}, &\qquad
\rho^a(\sigma,\tau)
  &= \sum_{n=-\infty}^\infty \rho^a_n(\tau) e^{-in\sigma}, \\
\varphi^\ib(\sigma,\tau)
  &=  s^{-1/2} \sum_{n \neq 0} \phi^\ib_n(\tau) e^{in\sigma}, &
\alpha^\ab(\sigma,\tau)
  &= \sum_{n=-\infty}^\infty \alpha_n^\ab(\tau) e^{-in\sigma}.
\end{alignedat}
\end{equation}
The factor $s^{-1/2}$ is introduced to keep track of the order of perturbation.  At the lowest order in perturbation theory, this is equivalent
to a Fourier expansion.

The bosonic left- and right-moving raising-lowering operators, denoted respectively by $c_n$ and $\tilde c_n$, are written in the Schr\"odinger
basis as \cite{Polchinski:1998rq}
\begin{equation}
\label{BRL}
\begin{aligned}
c_n^i
  &= -i\Bigl(g^{i\jb}(\phi_0) \frac{\del}{\del\phi_{-n}^\jb}
             + \frac{n}{2} \phi^i_n\Bigr), &\qquad
c_{n,i}
  &= -i\Bigl(\frac{\del}{\del\phi_{-n}^i}
             + \frac{n}{2} g_{i\jb}(\phi_0) \phi^\jb_n\Bigr), \\
\tilde c_n^i
  &= -i\Bigl(g^{i\jb}(\phi_0) \frac{\del}{\del\phi_n^\jb}
             + \frac{n}{2} \phi^i_{-n}\Bigr), &
\tilde c_{n,i}
  &= -i\Bigl(\frac{\del}{\del\phi_n^i}
             + \frac{n}{2} g_{i\jb}(\phi_0) \phi^\jb_{-n}\Bigr).
\end{aligned}
\end{equation}
They obey the familiar commutation relations $[c_m^i, c_{n,j}] = [\tilde c_m^i, \tilde c_{n,j}] = m\delta^i_j \delta_{m,-n}$.  The fermionic
modes, on the other hand, satisfy $\{\rho_{\ab,m}, \alpha^\bb_n\} = \delta_\ab^\bb \delta_{m,-n}$, where the zero mode part generates the
Clifford algebra.  Now, the Fock space is constructed by applying raising operators ($n < 0$) on a ground state given by a spinor $\psi$ on $X$:
\begin{equation}
\label{VAC} \ket{0;\psi} = \psi(\phi_0)
  \exp\Bigl(- \sum_{n = 1}^\infty \frac{n}{2}
            g_{i\jb}(\phi_0) \bigl(\phi_n^i\phi_{-n}^\jb
            + \phi_{-n}^i\phi_n^\jb \bigr)\Bigr)
  \prod_{n \geq 1} \prod_{\ab} \alpha_n^\ab,
\end{equation}
which is annihilated by lowering operators ($n > 0$), and on which $\rho_0$, $\alpha_0$ act by Clifford multiplication.  Every time a raising
operator is applied, it creates an index of either $TX$ or $T^*X$.  Thus, a general excited state is a spinor twisted by some combination of
tensor products of $TX$ and $T^*X$.

Expanding \eqref{QsEXPLICIT} at order $s^0$, we obtain the expressions for the supercharges in terms of the raising-lowering operators:
\begin{equation}
\begin{split}
Q_s &= \alpha_0^\ib \frac{D}{D\phi_0^\ib}
       + is^{1/2} \sum_{n \neq 0}
         g_{\ib j}(\phi_0) \alpha^\ib_{-n} \tilde c_n^j, \\
Q_s^\dagger
    &= -\rho_0^i \frac{D}{D\phi_0^i}
       - is^{1/2} \sum_{n \neq 0}
         g^{\ib j}(\phi_0) \rho_{-n,\ib} \tilde c_{n,j},
\end{split}
\end{equation}
where the covariant derivative with respect to $\phi_0$ is twisted by the bundle mentioned above.  The fermionic and the right-moving bosonic
raising operators increase $\Delta_s$ by order $s$; hence they do not enter the supersymmetric spectrum.  Restricted to states which have
$\Delta_s = 0$ at the lowest order, the supercharge $Q_s$ reduces to half of the twisted Dirac operator.  Therefore, we find that the
$Q$-cohomology, which is a spinor cohomology on $LX$, is approximated by a direct sum of certain twisted spinor cohomology groups on $X$.  More
precisely, states of the form $c_{-n_1}^{i_1} \dotsm c_{-n_k}^{i_k} \cdot c_{-\nb_1}^{\ib_1} \dotsm c_{-\nb_l}^{\ib_l} \ket{0;\psi}$ with $n_1 +
\dotsb + n_k = N$ and $\nb_1 + \dotsb + \nb_l = \Nb$ are sections of the bundle $S \otimes R_N \otimes R_\Nb^*$, where $\sum_{k = 0}^\infty q^k
R_k = \bigotimes_{k = 0}^\infty  \bigoplus_{l = 0}^\infty \Sym^l(q^k \cdot T^*X)$.

As long as the parameter $s$ is not strictly infinite, there are perturbative corrections to~$Q_s$.  For example, at the order $s^{-1/2}$, it
contains a term of the form
\begin{equation}
\alpha_0^\ib R_{\ib j\kb l}(\phi_0) \phi_{-2}^j \phi_1^\kb \phi_1^l \propto \alpha_0^\ib R_{\ib j\kb l}(\phi_0) (c_{-2}^j - \tilde c_2^j)
(c_1^\kb - \tilde c_{-1}^\kb)(c_1^l - \tilde c_{-1}^l).
\end{equation}
If we let this act on $\ket{T_{zz}} = g_{i\jb}(\phi_0) c_{-1}^i c_{-1}^\jb \ket{0;\psi}$, where $\psi$ is a harmonic spinor on~$X$, we obtain
\begin{equation}
\label{QTDELTHETA} \bigl(\alpha_0^\ib R_{\ib j\kb l}(\phi_0) c_{-2}^j c_1^\kb c_1^l\bigr) \cdot g_{i\jb}(\phi_0) c_{-1}^i c_{-1}^\jb
\ket{0;\psi} \propto R_{\ib j}(\phi_0) \alpha_0^\ib c_{-2}^j \ket{0;\psi} + \dotsb,
\end{equation}
where the abbreviated terms have nonzero values of $\Delta_s$ at the lowest order in perturbation theory.  Since quantum corrections can never
create new cohomology classes, we can immediately conclude the following.  The state $\ket{\del_z\theta} = R_{\ib j}(\phi_0) \alpha_0^\ib
c_{-2}^j \ket{0;\psi}$, which is $Q_s$-closed at the order $s^0$, gets perturbative corrections as shown on the right-hand side
of~\eqref{QTDELTHETA}.  However, $\ket{\del_z\theta}$ is now $Q_s$-exact, i.e., $Q_s\ket{\Psi} = \ket{\del_z\theta}$ for some $\ket{\Psi}$.
From~\eqref{QTDELTHETA}, we see that the state $\ket{\Psi}$ must be $\ket{T_{zz}}$ after the perturbative corrections are included.  We
recognize this as the Hilbert space counterpart of the operator relation $[Q, T_{zz}] = \del_z\theta$ in sigma model perturbation theory.

\subsection{Perturbative Spectrum of the $\CP^1$ Model}

Now that we have obtained a fairly clear picture of the perturbative supersymmetric spectrum, let us specialize to the case of our interest
where $X = \CP^1$.  To further simplify the analysis, we modify our function $h$.  Let $f$ be a Morse function on $X$, with discrete critical
points.  We then take $h$ to be
\begin{equation}
\label{hwithf} h = \int_A \gammah^*\omega + \frac{\nu}{2} \int_{S^1} d\sigma \gamma^* f,
\end{equation}
where $\nu$ is some real number.  As we will see shortly, this modification will ``push'' the perturbative supersymmetric states to the critical
points of $f$.

As before, the harmonic space in the large~$s$ limit is spanned by the critical points of~$h$, that are loops obeying the equation
\begin{equation}
g^{i\jb} \frac{\delta h}{\delta\phi^\jb} = i\del_\sigma\phi^i + \nu(\nabla f)^i = 0.
\end{equation}
One can easily show that $\del_\sigma f = 0$ for such configurations.  Although it is possible that the vector field $\nabla f$ generates a
$U(1)$ action, in which case one obtains nonconstant loops winding along contour lines of~$f$, one can kill these configurations by choosing a
noninteger value for~$\nu$.  Then, only constant loops at the critical points of~$f$ will be left.  We will take $0 < \nu < 1$ in what follows.

Next, we cover $X=\CP^1$ with two open sets $U_S = \CP^1 \setminus \{N\}$ and $U_N = \CP^1 \setminus \{S\}$, described by coordinates $\phi$ and
$\phi' = 1/\phi$, respectively.  We equip the target space with the Fubini-Study metric
\begin{equation}
g(\phi,\phib) = \frac{1}{(1 + |\phi|^2)^2}
  (d\phi\otimes d\phib + d\phib\otimes d\phi).
\end{equation}
For the Morse function $f$, we use
\begin{equation}
f(\phi,\phib) = \frac{1}{2} \frac{|\phi|^2 + 1}{|\phi|^2 - 1},
\end{equation}
which generates the vector field $\nabla f = \phi\del_\phi + \phib\del_\phib = -\phi'\del_{\phi'} - \phib'\del_{\phib'}$.  The critical points
of~$f$ are thus~$S$ with $\phi = 0$ and~$N$ with $\phi' = 0$.  Localized around each critical point are perturbative supersymmetric states.  We
now identify the spaces $\CH_S$ and $\CH_N$ formed by these states.

Let us go back for a moment to the two supercharges deformed by the function~\eqref{MLX}.  In this case, perturbative supersymmetric states are
twisted spinors, whose supports in general spread over the whole target space.  The perturbative supersymmetric spectrum naturally splits into
two subspaces as $\CH_0 \oplus \CH_1$ according to the fermionic number, or equivalently, the chirality of states.  We first consider the space
$\CH_0$.  Perturbatively, states with fermionic number~$0$ are trivially annihilated by $Q_s^\dagger$.  Then, states in $\CH_0$ are
characterized by the property that they are annihilated by $Q_s$.  Now, the supercharge deformed by the function~\eqref{hwithf} can be obtained
by further taking the conjugate $Q_s \to e^{s\nu f/2\pi} Q_s e^{-s\nu f/2\pi}$.  Accordingly, $\CH_0$ is mapped to an isomorphic space by
$\ket{\Psi_0} \to e^{s\nu f/2\pi} \ket{\Psi_0}$.  States with fermionic number~$0$ are then pushed in the limit $s \to \infty$ to~$S$ where~$f$
is minimum.  Thus, we find that $\CH_S \iso \CH_0$.  By exchanging the roles of $Q_s$ and $Q_s^\dagger$ in the above argument, we see that
states with fermionic number~$1$ are pushed to~$N$ where~$f$ is maximum, and we have $\CH_N \iso \CH_1$.  It is straightforward to generalize
this construction to target spaces of higher dimensions.  In general, the perturbative supersymmetric states localized at a critical point with
Morse index $p$ have fermionic number~$p$.

The spaces $\CH_0$ and $\CH_1$ are isomorphic to one another; they are related by a complex conjugation on the target space.  We then also have
$\CH_S \iso \CH_N$.  This implies that instantons can pair up all the states in $\CH_S$ and $\CH_N$, and lift them from the supersymmetric
spectrum.

\subsection{Instantons in the $\CP^1$ Model}

We have described above the perturbative supersymmetric spectrum of the $\CP^1$ model.  Starting from this space, we can compute explicitly the
full supersymmetric spectrum by considering the Witten complex~\cite{Witten:1982im}, in which the nonperturbative action of $Q_s$ is given by
the quantum tunneling between states localized at different critical points.

A nonzero matrix element $\bra{\Psi_f} Q_s \ket{\Psi_i}$ indicates the existence of instantons connecting the initial state $\Psi_i$ and the
final state $\Psi_f$.  Note here that $\Psi_i$ can have positive energy, since supersymmetric states in the $\CN = (0,2)$ case need only to
satisfy $\{Q,Q^\dagger\} = (H - P)/2 = 0$ (or a similar condition deformed by~$h$).  If this is the case, $\Psi_i$ decays exponentially fast as
it propagates and will hence never reach the other end of the infinitely long cylinder.  Thus, we must use $e^{-(H - P)\tau}$ in place of
$e^{-H\tau}$ to propagate this state along the cylinder, canceling the decay by the exponential factor $e^{P\tau}$.  Using the deformed
supercharges, we can express the matrix element as
\begin{equation}
\label{AMP} \bra{\Psi_f} Q_s \ket{\Psi_i} = \lim_{T \to \infty} \frac{1}{h(\gamma_T) - h(\gamma_{-T})} \bra{\Psi_f} e^{-2\Delta_s T} [Q_s,h]
e^{-2\Delta_s T} \ket{\Psi_i}
 + \CO(s^{-1/2}),
\end{equation}
where $\gamma_\tau = \phi(\blank,\tau) \in LX$  (see \cite{Hori:2003ic}, Chapter~10).  A straightforward computation shows that $\Delta_s = (H -
sP_\nu)/2$, where $P_\nu = P + i\nu J$ and $J$ is the complex structure of~$X$.

The amplitude that appears on the right-hand side of \eqref{AMP} can be computed via the path integral around instantons:
\begin{equation}
\label{AMP-PI} \int \CD\phi\CD\rho\CD\alpha \, e^{-S_h + 2sP_\nu(\Psi_i)T} \Psi_f^\dagger(\gamma_T) \Bigl(\int_{S^1} d\sigma \alpha^\ib
\frac{\delta h}{\delta \phi^\ib}\Bigr) \Psi_i(\gamma_{-T}).
\end{equation}
One obtains the equation which governs instanton configurations by setting the $Q_s$-variation of the fermions to zero: $\del_\tau\phi^i -
2sg^{i\jb} \delta h/\delta\phi^\jb = 0$.  Thus, instantons are ascending gradient flows of the Morse function~$h$ on~$LX$.  Explicitly, it reads
\begin{equation}
(\del_\tau - is\del_\sigma - s\nu) \phi = (\del_\tau - is\del_\sigma + s\nu) \phi' = 0.
\end{equation}
The wavefunctions $\Psi_i$, $\Psi_f$ are sharply peaked at~$S$ or~$N$.  Appropriately normalized, they contain delta functions centered at the corresponding poles in the limit $s \to \infty$.  Thus, we have only to look for instantons connecting the two poles.  Recalling that $0 < \nu < 1$, we see that instantons going from $S$ to $N$ are $\phi \propto e^{-in\sigma + s(n + \nu)\tau}$ for $n \geq 0$, while those going from $N$ to $S$ are $\phi' \propto e^{-in\sigma + s(n - \nu)\tau}$ for~$n \geq 1$.  The path integral \eqref{AMP-PI} vanishes unless we have precisely one $\alpha$ zero mode and no $\rho$ zero modes.  This condition leaves%
\footnote{The equation for $\alpha$ zero modes is $(\del_\tau + is\del_\sigma - s\nu)\ub^\ib = 0$, and that for $\rho$ zero modes is $(\del_\tau
+ is\del_\sigma + s\nu) v_i = 0$, written in the coordinates of $U_S$.  They are solved by $\ub \propto e^{il\sigma + s(l + \nu)\tau}$ and $v
\propto e^{im\sigma + s(m - \nu)\tau}$.  Imposing $\|\ub\|$, $\|v\| \to 0$ as $\tau \to \pm\infty$, one obtains the conditions $-\nu < l < 2n +
\nu$ and $\nu < m < -2n - \nu$ for instantons going from~$S$ to~$N$.  Similarly, one finds $\nu < l < 2n - \nu$ and $-\nu < m < -2n + \nu$ for
instantons going from~$N$ to~$S$.}
only $\phi \propto e^{s\nu\tau}$ and $\phi' \propto e^{-i\sigma + s(1 - \nu)\tau}$.  The same conclusion can be obtained by requiring that the
Morse index of~$h$, which is equal to the fermionic number, increases by one as instantons goes from one critical point to another.

As a first example of an instanton effect, we consider the amplitude going from a ground state $\ket{0;S}$ at~$S$ to another ground state
$\ket{0;N}$ at~$N$.  These states are connected by a family of instantons
\begin{equation}
\phi_{\sigma_0, \tau_0}(\tau) = e^{s\nu(\tau - \tau_0) + i\sigma_0}
\end{equation}
parameterized by $\sigma_0$, $\tau_0 \in \R$.  Since these are worldline instantons which do not depend on $\sigma$, the problem reduces to
supersymmetric quantum mechanics on~$X$.  The action~\eqref{S_h} for these instanton configurations is, as~$h$ in the case of worldline
instantons is simply given by~$f$,
\begin{equation}
S_h = \frac{s}{2\pi} \int_{-\infty}^\infty d\tau
      \del_\tau h(\phi_{\sigma_0, \tau_0})
    = \frac{s}{2\pi} \bigl(f(N) - f(S)\bigr).
\end{equation}
After the integration over the fermionic zero mode is done, we perform the integration over the instanton moduli space and obtain, up to a
numerical factor,
\begin{equation}
\begin{split}
\int d\tau_0 d\sigma_0 \, e^{-S_h} \int_{S^1} d\sigma \del_{\tau_0} \phib_{\sigma_0, \tau_0} \frac{\delta h}{\delta \phib}(\phi_{\sigma_0,
\tau_0}) &\propto
   \int d\tau_0 d\sigma_0 \, e^{-S_h}
   \del_{\tau_0} h(\phi_{\sigma_0, \tau_0}) \\
& \propto
   \bigl(f(N) - f(S)\bigr) e^{-s(f(N) - f(S))/2\pi}.
\end{split}
\end{equation}
Here, we have used the fact that neither the wavefunctions nor the Morse function~$h$ depend on the phase of~$\phi_{\sigma_0, \tau_0}$, and
$\phi_{\sigma_0, \tau_0} = \phib_{\sigma_0, \tau_0}$ up to a phase.  Plugging the above result in~\eqref{AMP}, we finally obtain
\begin{equation}
\bra{0;N} Q_s \ket{0;S} \propto (\Det'\Delta_B)^{1/2} e^{-s(f(N) - f(S))/2\pi},
\end{equation}
where $\Det'\Delta_B$ is the bosonic determinant that is only partially cancelled by the fermionic determinant since the theory lacks
left-moving fermions.  Since instantons are ascending gradient flows of~$h$, this amplitude is exponentially suppressed, indicating that it
arises from nonperturbative effects.

Since there is no other destination which $Q_s\ket{0;S}$ can connect to, we conclude that $\ket{0;N} \propto Q_s\ket{0;S}$.  Note that this
relation conserves the fermionic number.  This is because worldline instantons are ``classical'' instantons, which detect the classical geometry
of the target: As $\ket{0;S}$ and $\ket{0;N}$ are ground states corresponding to harmonic spinors, this phenomenon is just a reflection of
Lichnerowicz's theorem in classical geometry that there are no harmonic spinors on $\CP^1$ because of its positive scalar curvature.

The second example is provided by $\ket{\Psi_i} = c_{-1}\ket{0;N}$ and $\ket{\Psi_f} = \cb_{-1} \ket{0;S}$.  From the expressions~\eqref{BRL}
and~\eqref{VAC} in the Schr\"odinger basis, we see that $\ket{\Psi_i} \propto \phi_{-1}'\ket{0;N}$ and $\ket{\Psi_f} \propto \phib_{-1}
\ket{0;S}$.  Since the instantons
\begin{equation}
\phi'_{\sigma_0, \tau_0}(\sigma, \tau) = e^{-i(\sigma - \sigma_0) + (1-\nu)s(\tau - \tau_0)}
\end{equation}
have excitations of the modes $\phi'_{-1}$ and $\phib_{-1}$, they may connect the initial and final states.  Although perturbatively
$Q_s\ket{\Psi_i}$ has fermionic number~$2$ and~$\ket{\Psi_f}$ has~$0$, this lifting is possible due to the charge anomaly---the charge violation
is exactly~$\phi^*c_1(X) = 2$ for instantons of degree one on~$X = \CP^1$.  This time the path integral~\eqref{AMP-PI} contains the factor $
\phi_{1}(T) \phi'_{-1}(-T) \propto e^{-2s(1 - \nu)T} $ from the wavefunctions, but this is cancelled by $e^{2sP_\nu(\Phi_i)T}$.  The rest of the
computation proceeds as in the first example, and we obtain
\begin{equation}
\bra{\Psi_f} Q_s \ket{\Psi_i} \propto (\Det'\Delta_B)^{1/2}
        e^{-s(h(\gamma_\infty) - h(\gamma_{-\infty}))/2\pi}.
\end{equation}
In contrast to the first example, this nonvanishing amplitude is induced by worldsheet instantons which depend on both $\sigma$ and $\tau$.  They wrap the target space once as they propagate from $N$ to $S$, sweeping out its fundamental class.  There is an interesting complication in determining the quantity $h(\gamma_\infty) - h(\gamma_{-\infty})$, due to the multivaluedness of the function~$h$.%
\footnote{To see why it arises, we must go back to the definition of~$h$.  Let us look at the part given by the K\"ahler form of~$X$.  The loop
space $LX$ is connected for $X = \CP^1$, so we can take the constant loop at~$S$ to be our reference point~$\gamma_0 \in LX$.  Given another
point~$\gamma \in LX$, we construct a homotopy~$\gamma_u$ connecting~$\gamma_0$ and $\gamma_1 = \gamma$.  In principle, there are many different
choices for this.  Any two such homotopies can be combined to a closed path in~$LX$ that defines a map $S^2 \to X$, where the two poles of the
sphere are identified.  If this defines a nontrivial $2$-cycle in the target, then the value of~$h$ changes by the K\"ahler class evaluated on
that cycle.}
However, in the present case we have instantons connecting two critical points on~$LX$ and they are all homotopically equivalent.  Therefore,
the \emph{relative} value of $h$ is still well defined.  For the worldsheet instantons, this is given by the usual weight for instantons of
degree one, and the contribution from the Morse function~$f$ on~$X$.

There is also a physically intuitive way to understand why the above lifting is possible.  Since the connection vanishes at~$N$ and~$S$, the
covariant derivative~$iD_\sigma$ reduces to~$i\del_\sigma$ at $\tau = \pm\infty$.   Suppose that we have an eigenmode $\vphib_n(\sigma,\tau)$ of
$iD_\sigma$, whose eigenvalue at $\tau = -\infty$ is~$n$, so that $\vphib_n(\sigma,-\infty) = e^{-in\sigma}$.  Let us see what happens to the
eigenvalue as this eigenmode propagates along a worldsheet instanton configuration.  Letting~$iD_\sigma$ act on~$\vphib_n$, we have
\begin{equation}
\label{iDsz}
\begin{split}
iD_\sigma\vphib_n = i\del_\sigma \vphib_n
   + i\del_\sigma \phib' \Gamma^{\phib'}_{\phib'\phib'} \vphib_n
= i\del_\sigma \vphib_n + \frac{2|\phi'|^2}{1 + |\phi'|^2} \vphib_n
\end{split}
\end{equation}
in the coordinate patch~$U_N$.  The $\sigma$-dependence of~$\vphib_n$ remains to be~$e^{-in\sigma}$, since the eigenvalues of~$i\del_\sigma$ can
only take integer values.  On the other hand, the connection term changes from~$0$ to~$2$ as the instanton propagates from~$N$ to~$S$.  We then
see that the eigenmode~$\vphib_n$ must have eigenvalue $n+2$ at $\tau = +\infty$, which means that $\vphib_n(\phi,+\infty) \propto
e^{-i(n+2)\sigma}$ in the coordinates of~$U_S$.  Thus, we have found the spectral flow $\alpha_n \to \alpha_{n+2}$: As the state
$c_{-1}\ket{0;N}$ propagates from~$N$ to~$S$, the structure of the Dirac sea is altered as
\begin{equation}
\prod_{n = 0}^\infty \alpha_n \longto \prod_{n = 2}^\infty \alpha_n.
\end{equation}
Thus, there is now a hole in the Dirac sea.  This can be filled by $\alpha_1$ contained in $Q_s$, and one obtains the Dirac sea for states
localized at~$S$, which do not have~$\alpha_0$.

\subsection{Loop Space and Quantum Geometry}

Let us summarize what we have found in this section.  We have seen, by deforming the supercharges using a Morse function, that the
$Q$-cohomology of states for $X = \CP^1$ is given perturbatively by the Fock spaces of closed strings with fermionic number~$0$ and~$1$ located
respectively at~$S$ and~$N$.  States are therefore constructed from perturbative supersymmetric ground states~$\ket{0;S}$ and~$\ket{0;N}$.  At
this level, the bosonic and fermionic supersymmetric spectra are manifestly isomorphic due to the double degeneracy of the Ramond sector.  The
classical geometry of $\CP^1$ induces a ``classical'' instanton effect that lifts~$\ket{0;S}$ and~$\ket{0;N}$ from the supersymmetric spectrum.
In addition, worldsheet instantons lift~$c_{-1} \ket{0;N}$ via the relation $Q c_{-1} \ket{0;N} \sim \cb_{-1} \ket{0;S} + \dotsb$.  The same
argument also applies to excited supersymmetric states.  In this case~$H$ and~$P$ can take arbitrary value satisfying the condition $H = P$, and
the amplitude~\eqref{AMP} can be nonvanishing since we are using the operator $e^{-(H - P)\tau}$ instead of $e^{-H\tau}$ to propagate the states
along the cylinder.

Since supersymmetry is spontaneously broken, all excited states will be lifted either by the worldline or worldsheet instantons described above,
or by perturbative corrections to~$Q$.  To ascertain explicitly how each and every state is lifted is challenging, at least from the viewpoint
of Morse theory on loop space.  Nevertheless, some of the cases can be easily described.  For example, states of the form $V_i c_{-n}^i
\ket{0;S}$ and $V^i c_{-n,i} \ket{0;N}$ are sections of the bundle $S\otimes T^*X \iso \CO(-3)$ and $S\otimes TX \iso \CO(1)$, respectively
(before the deformation by~$f$).  Since $h^0(\CO(-3)) = h^1(\CO(1)) = 0$, these states must be lifted by classical, worldline instantons.  Thus,
states of the former type are not $Q$-closed, and those of the latter type are all $Q$-exact.  On the other hand, states of the form $V^i
c_{-n,i} \ket{0;S}$ and $V_i c_{-n}^i \ket{0;N}$ are sections of the bundle $S\otimes TX \iso \CO(1)$ and $S\otimes T^*X \iso \CO(-3)$,
respectively, and the Betti numbers are~$2$ for both cases.  For $n = 1$, we saw above that worldsheet instantons lift them.  For $n \geq 2$,
worldsheet instantons cannot lift them, but perturbative corrections will destroy the classes.  We have seen one such mechanism at the end of
Section~\ref{PSS}, which led to the relation $Q\ket{T_{zz}} = \ket{\del_z\theta}$.

Since our model is formulated as supersymmetric quantum mechanics on the loop space~$LX$, worldsheet instantons will detect the classical
geometry of~$LX$; they are analogous to worldline instantons---which are instantons in supersymmetric quantum mechanics on~$X$---that detect the
classical geometry of~$X$.  Our results tell us that the $Q$-cohomology of $X = \CP^1$, i.e., a spinor cohomology of~$LX$, is trivial.  This is
consistent with Stolz's idea~\cite{S} that if~$X$ has positive Ricci curvature, then $LX$ will have a positive Ricci scalar with no harmonic
spinors.  From the target space viewpoint, one could say that under a ``quantum'' deformation of the geometry of~$\CP^1$ by worldsheet
instantons, the kernels of the corresponding twisted Dirac operators become empty.

\section*{Acknowledgements}

The authors would like to thank E.  Frenkel, S.  Katz, E.  Sharpe, S.  Stolz and most of all A.~Basu and E.  Witten, for illuminating and
helpful discussions.  J.Y.  would also like to thank G.~Moore for his inspiring lectures.  The work of M.-C.T.  is supported by the Institute
for Advanced Study and the NUS -- Overseas Postdoctoral Fellowship.

\end{document}